\documentclass[a4paper,11pt]{article}
\pdfoutput=1 

\usepackage{jheppub} 
\usepackage{physics}

\usepackage{tikz}
\usetikzlibrary{calc}

\usepackage[T1]{fontenc} 

\newcommand\fft[2]{\frac{#1}{#2}}

\newcommand\nn{\nonumber}

\preprint{LCTP-18-12}

\title{\boldmath The topologically twisted index of $\mathcal N=4$ super-Yang-Mills on $T^2\times S^2$ and the elliptic genus}

\author{Junho Hong,}
\author{James T. Liu}

\affiliation{Leinweber Center for Theoretical Physics, Randall Laboratory of Physics \\The University of Michigan, Ann Arbor, MI 48109-1040, USA }

\emailAdd{junhoh@umich.edu}
\emailAdd{jimliu@umich.edu}

\makeatletter
\newcommand*{\rom}[1]{\expandafter\@slowromancap\romannumeral #1@}
\makeatother

\abstract{We examine the topologically twisted index of $\mathcal N=4$ super-Yang-Mills with gauge group $SU(N)$ on $T^2\times S^2$, and demonstrate that it receives contributions from multiple sectors corresponding to the freely acting orbifolds $T^2/\mathbb Z_m\times\mathbb Z_n$ where $N=mn$.  After summing over these sectors, the index can be expressed as the elliptic genus of a two-dimensional $\mathcal N=(0,2)$ theory resulting from Kaluza-Klein reduction on $S^2$.  This provides an alternate path to the `high-temperature' limit of the index, and confirms the connection to the right-moving central charge of the $\mathcal N=(0,2)$ theory.}

\begin{document} 
\maketitle
\flushbottom

\section{Introduction}

Recent advances in supersymmetric field theories have led to a new era of precision holography through AdS/CFT. It has been driven on the field theory side by the key developments of rigid supersymmetry \cite{Festuccia:2011ws} and supersymmetric localization \cite{Pestun:2007rz}. The former allows for a systematic construction of supersymmetric field theories on a curved background with topological twists. The latter yields exact field theory results reliable even at strong coupling limit. Combining these two developments, now we can compute exact field theory results of various topologically twisted SCFTs on curved backgrounds, which can be explored in the holographic dual through AdS/CFT. Of particular interest are partition functions on $S^d$ which gives the free energy and on $S^1\times S^{d-1}$ which computes the supersymmetric index as well as Wilson loop observables in various representations of the gauge group.  

In particular, a three-dimensional topologically twisted index was introduced as the supersymmetric index on $S^1\times S^2$ with a topological twist on $S^2$ \cite{Benini:2015noa}. When applied to the ABJM theory \cite{Aharony:2008ug}, it has an interesting feature. In the large-$N$ limit, the topologically twisted index of the ABJM theory on $S^1\times S^2$ matches the entropy of the dual asymptotically AdS$_4$ magnetic black hole, when it is extremized with respect to the chemical potentials \cite{Benini:2015eyy}. This is regarded as the first counting of the microstates of a supersymmetric asymptotically AdS$_4$ black hole.

Similarly, the four-dimensional topologically twisted index can be introduced as the supersymmetric index on $T^2\times S^2$ with a topological twist on $S^2$. In particular, we can apply this to $\mathcal N=4~SU(N)$ super-Yang-Mills (SYM) with a similar goal in mind, namely counting the microstates of the dual asymptotically AdS$_5$ magnetic black string. This is still an open problem, however, and here we review some of the unsolved issues in both the field theory and supergravity sides of the duality.

{\bf Field theory side:} The topologically twisted $\mathcal N=4~SU(N)$ SYM on $T^2\times S^2$ can be constructed by equipping $S^2$ with background gauge fields that couple to the $SO(6)$ R-symmetry current, satisfying the conditions categorized in \cite{Benini:2013cda}. The explicit computation of the topologically twisted index in the large $N$ limit, however, has not yet been performed unlike in the ABJM theory case. Instead, it has been investigated in the `high-temperature' limit, $\beta\rightarrow0^+$, where the modular parameter of the torus is given by $\tau={i\beta}/{2\pi}$ \cite{Hosseini:2016cyf}.

{\bf Supergravity side:} The holographic dual of $\mathcal N=4~SU(N)$ SYM on $T^2\times S^2$ has been studied in \cite{Maldacena:2000mw,Benini:2013cda}. To be specific, based on the well known duality between $\mathcal N=4~SU(N)$ SYM on $\mathbb R^{1,3}$ and Type \rom{2}B supergravity in AdS$_5\times S^5$ background, we may expect that the same field theory on $T^2\times S^2$ with topological twists is holographically dual to Type \rom{2}B supergravity in an asymptotically AdS$_5$ magnetic black string background with conformal boundary $T^2\times S^2$.  The AdS$_3\times S^2$ near-horizon solution for the string is known and numerical evidence suggests that it can be extended into a full solution \cite{Benini:2013cda}.  However, a full analytic supergravity solution with such asymptotic conditions has yet to be constructed%
\footnote{Asymptotically AdS$_5$ black hole solutions with conformal boundary $\mathbb R\times S^3$ have been constructed in \cite{Gutowski:2004ez,Gutowski:2004yv}. Even in this case, however, matching its entropy with microstate counting in the large-$N$ limit of the dual field theory has not yet been done due to various issues.}.

At this stage, we focus on the field theory side by taking a closer look at the topologically twisted index of $\mathcal N=4~SU(N)$ SYM on $T^2\times S^2$. As demonstrated in \cite{Hosseini:2016cyf}, the high-temperature limit of the index, when extremized over the chemical potentials $\Delta_a$, matches the right-moving central charge of the $\mathcal N=(0,2)$ SCFT associated with the AdS$_5$ magnetic black string
\begin{equation}
    \left.\Re\log Z\left(\tau=\fft{i\beta}{2\pi},\bar\Delta_a,\mathfrak n_a\right)\right|_{\beta\to0^+}=\fft{\pi^2}{6\beta}c_r(\mathfrak n_a),
    \label{eq:Cardy}
\end{equation}
where $\{\mathfrak n_a\}$ are integer magnetic charges satisfying $\sum_{a=1}^3\mathfrak n_a=2$ \cite{Hosseini:2016cyf}, two of them being negative \cite{Benini:2013cda}.  In a way, this is not surprising, as (\ref{eq:Cardy}) is just the expected behavior in the Cardy limit of the SCFT.  Away from this limit, however, the index must transform as a weak Jacobi form.  This can be seen by Kaluza-Klein reducing on $S^2$, whereupon the supersymmetric index on $T^2$ becomes the elliptic genus \cite{Honda:2015yha}.

In this paper, we clarify the connection between the topologically twisted index of $\mathcal N=4$, $SU(N)$ SYM on $T^2\times S^2$ and the elliptic genus.  As constructed in \cite{Hosseini:2016cyf}, the index can be computed using Jeffrey-Kirwan residues.  The result is thus given in terms of a sum over solutions to a set of algebraic equations, commonly referred to as the `Bethe ansatz equations' (BAEs).  In contrast to the $S^2\times S^1$ index, where there is only a single solution to the BAEs (up to permutations) \cite{Benini:2015noa,Benini:2015eyy}, here we find multiple solutions, where the `eigenvalues' are uniformly distributed over the $T^2$.  Furthermore, the existence of these multiple solutions is fundamental in order for the index to be an elliptic genus.

Once the index is understood as an elliptic genus, we revisit the high-temperature limit, $\tau\to i0^+$, by performing the modular transformation $\tau\to-1/\tau$. Although some of our results are left at the conjecture level, we reproduce the Cardy limit (\ref{eq:Cardy}), where
\begin{equation}
    c_r(\mathfrak n_a)=3(N^2-1)\fft{\mathfrak n_1\mathfrak n_2\mathfrak n_3}{1-(\mathfrak n_1\mathfrak n_2+\mathfrak n_2\mathfrak n_3+\mathfrak n_3\mathfrak n_1)},
    \label{central:charge}
\end{equation}
in agreement with \cite{Hosseini:2016cyf}.  Since this expression is valid for arbitrary $N$, it also holds in the large-$N$ case with holographic dual.  More generally, however, it would be interesting to explore the large-$N$ limit at arbitrary values of the modular parameter $\tau$.  Unfortunately, this still appears to be a rather challenging problem, as the only expression we have for the index at arbitrary $N$ is given as a sum over sectors, each corresponding to a different solution to the BAEs.

The outline of the paper is as follows.  In section~\ref{sec:2}, we first review the topologically twisted index of $\mathcal N=4$ SYM on $T^2\times S^2$, then demonstrate that the BAEs admit multiple solutions.  In section~\ref{sec:3}, we connect the index to the elliptic genus and in particular demonstrate that it transforms as a weak Jacobi form. Given this understanding of its modular properties, we then revisit the high-temperature limit in section~\ref{sec:4}.  Finally, we conclude with some comments on the large-$N$ limit in section~\ref{sec:5}.

\section{The topologically twisted index of $\mathcal N=4$ SYM on $T^2\times S^2$}
\label{sec:2}

The topologically twisted index of $\mathcal N=4$ SYM with gauge group $SU(N)$ was defined in \cite{Benini:2015noa,Hosseini:2016cyf} as the supersymmetric index of the theory on $T^2\times S^2$ with a topological twist on $S^2$. The index depends on the modular parameter $q=e^{2\pi i\tau}$ as well as flavor chemical potentials $\Delta_a$ and magnetic fluxes $\mathfrak n_a$, and may be written as \cite{Hosseini:2016cyf}
\begin{equation}\label{generating functional}
    Z(\tau;\Delta_a,\mathfrak n_a)=\mathcal{A}\sum_{I\in \mathrm{BAEs}}\frac{1}{\det\mathbb B}\prod_{j\neq k}^N\prod_{a=1}^3\left(\frac{\theta_1(u_j-u_k;\tau)}{\theta_1(u_j-u_k+\Delta_a;\tau)}\right)^{1-\mathfrak n_a},
\end{equation}
where the prefactor $\mathcal A$ is given by
\begin{equation}
    \mathcal{A}=i^{N-1}\eta(\tau)^{3(N-1)}\prod_{a=1}^3\theta_1(\Delta_a;\tau)^{-(N-1)(1-\mathfrak n_a)}.
\label{eq:Afactor}
\end{equation}
Definitions and modular properties of the Dedekind eta function $\eta$ and the Jacobi theta function $\theta_1$ are given in Appendix~\ref{App A}.   

The sum in (\ref{generating functional}) is over all solutions, $I=\{u_0,u_1,\ldots,u_{N-1},v\}$, of the `Bethe ansatz equations' (BAEs)
\begin{equation}\label{BAEs}
    e^{iB_j}=1\qquad(j\in\{0,1,\cdots,N-1\}),
\end{equation}
where
\begin{equation}
    B_j\equiv v+i\sum_{k=0}^{N-1}\sum_{a=1}^3\log\left(\frac{\theta_1(u_k-u_j+\Delta_a;\tau)}{\theta_1(u_j-u_k+\Delta_a;\tau)}\right).
    \label{B_j}
\end{equation}
Note that the $u_j$'s are also constrained to satisfy the $SU(N)$ condition $\sum_{j=0}^{N-1} u_j=0$.  In terms of (\ref{B_j}), the $N\times N$ Jacobian matrix $\mathbb B$ takes the form
\begin{equation}
    \mathbb B\equiv \frac{\partial(B_1,\cdots,B_N)}{\partial(u_1,\cdots,u_{N-1},v)}.
\label{B matrix}
\end{equation}
While this Jacobian is explicitly constructed from $N-1$ of the $N$ eigenvalues $u_i$, it is easily seen that it does not depend on which one is omitted because of the $SU(N)$ condition that the $u_i$'s sum to zero.

According to \cite{Hosseini:2016cyf}, the flavor chemical potentials $\Delta_a$ and the magnetic fluxes $\mathfrak n_a$ are constrained to satisfy
\begin{equation}
	\sum_{a=1}^3\Delta_a=2\pi\mathbb Z\qquad\mathrm{and}\qquad\sum_{a=1}^3\mathfrak n_a=2.
	\label{constraints}
\end{equation}
Here we exclude $\Delta_a=2\pi\mathbb Z$ in order to avoid issues with the vanishing of $\theta_1(0;\tau)$ but we do not necessarily assume $\Delta_a\in\mathbb R$ or $0<\Re\Delta_a<2\pi$. Instead, the twisted index (\ref{generating functional}) is invariant under
\begin{equation}
	\Delta_a\to-\Delta_a\qquad\mbox{and}\qquad\Delta_a\to\Delta_a+2\pi\mathbb Z
	\label{Deg:free}
\end{equation}
up to sign and we will fix these degrees of freedom later according to our purpose.

\subsection{Multiple solutions to the BAEs}

A solution to the BAEs, (\ref{BAEs}), was obtained in the `high-temperature' limit, $\tau\to i0^+$, in \cite{Hosseini:2016cyf} under the condition $\sum_{a=1}^3\Delta_a=2\pi$. It can be written as
\begin{equation}
    u_j=\bar u-2\pi\tilde\tau j,\qquad
    v=(N+1)\pi,
\label{First:solution}
\end{equation}
where $\bar u$ is a constant chosen to enforce the $SU(N)$ condition $\sum_j u_j=0$, and $\tilde\tau=\tau/N$. While this solution was obtained in the high-temperature limit, it actually satisfies the BAEs for any $\tau$ in the upper half plane and for arbitrary $\Delta_a$'s satisfying the constraint $\sum_{a=1}^3\Delta_a=2\pi\mathbb Z$.

Furthermore, we show below that (\ref{First:solution}) is in fact a special case of a larger set of BAE solutions. The key observation is that the $u_j$ variables are doubly periodic, as they are defined on $T^2$, with periods $u_j\to u_j+2\pi$ and $u_j\to u_j+2\pi\tau$. Based on this periodicity, the solution (\ref{First:solution}) then corresponds to the $u_j$'s being evenly distributed along the thermal circle. This defines the torus $T^2/\mathbb Z_N$ with modular parameter $\tilde\tau=\tau/N$. Then modular invariance suggests that having the $u_j$'s evenly distributed along the other cycle of the $T^2$ ought to yield another solution, this time with modular parameter $\tilde\tau=N\tau$. 

Taking this one step further, we expect that any $u_j$'s evenly distributed over the torus $T^2$ satisfy the BAEs, (\ref{BAEs}). In this case, the set of $u_j$'s defines a freely acting orbifold $T^2/\mathbb Z_m\times\mathbb Z_n$ where $\{m\}$ is the set of all positive divisors of $N$ with $N=mn$. The corresponding $u_j$'s can be written explicitly as
\begin{equation}
	u_{\hat j\hat k}=\bar u+2\pi\left(\fft{\hat j}m+\fft{\hat k}n\left(\tau+\fft{r}m\right)\right)=\bar u+2\pi\fft{\hat j+\hat k\tilde\tau}m\quad\mbox{where}\quad\tilde\tau\equiv\fft{m\tau+r}{n}.
	\label{eq:full BAE}
\end{equation}
Note that we have introduced a double index notation
\begin{equation}
	u_j\equiv u_{\hat j\hat k},\qquad \hat j=0,\ldots,m-1,\qquad\hat k=0,\ldots,n-1,
\end{equation}
and $r=0,\ldots,n-1$ is a constant that, along with $m$ and $n$, specifies the orbifold.

In order to prove that (\ref{eq:full BAE}) indeed satisfies the BAEs, we substitute it into (\ref{B_j}), so that the BAEs reduce to the claim that
\begin{equation}\label{BAE:orbifolds}
e^{iv}\overset{!}{=}\prod_{a=1}^3\prod_{\hat j=0}^{m-1}\prod_{\hat k=0}^{n-1}\fft{\theta_1\left(\Delta_a-2\pi\fft{(\hat j-\hat j_0)+(\hat k-\hat k_0)\tilde\tau}m;\tau\right)}{\theta_1\left(\Delta_a+2\pi\fft{(\hat j-\hat j_0)+(\hat k-\hat k_0)\tilde\tau}m;\tau\right)}.
\end{equation}
We now use the double periodicity of $\theta_1$, (\ref{eq:uperiod}), to shift the product over $\hat j$ and $\hat k$ as
\begin{align}\label{shifting}
	&\prod_{\hat j=-\hat j_0}^{m-\hat j_0-1}\prod_{\hat k=-\hat k_0}^{n-\hat k_0-1}\theta_1\left(\Delta_a\pm 2\pi\fft{\hat j+\hat k\tilde\tau}m;\tau\right)\nn\\&=(-1)^{n\hat j_0+(r-1)\hat k_0}e^{\pm i\hat k_0(m\Delta_a\pm(2n-\hat k_0-1)\pi\tilde\tau)}e^{-i\pi m\hat k_0\tau}\prod_{\hat j=0}^{m-1}\prod_{\hat k=0}^{n-1}\theta_1\left(\Delta_a\pm 2\pi\fft{\hat j+\hat k\tilde\tau}m;\tau\right).
\end{align}
Inserting this into the RHS of (\ref{BAE:orbifolds}) and using the constraint $\sum_{a=1}^3\Delta_a=2\pi\mathbb Z$, then gives
\begin{equation}
	e^{iv}\overset{!}{=}\prod_{a=1}^3\prod_{\hat j=0}^{m-1}\prod_{\hat k=0}^{n-1}\fft{\theta_1\left(\Delta_a-2\pi\fft{\hat j+\hat k\tilde\tau}m;\tau\right)}{\theta_1\left(\Delta_a+2\pi\fft{\hat j+\hat k\tilde\tau}m;\tau\right)}.
	\label{After:shifting}
\end{equation}
In particular, the RHS is now manifestly independent of $\hat j_0$ and $\hat k_0$, thus demonstrating that the full set of BAEs reduce to a single equation that can be consistently satisfied for an appropriately chosen $v$.

While this is sufficient to demonstrate that (\ref{eq:full BAE}) satisfies the BAEs, we can explicitly determine $v$ by choosing $\hat j_0=m-1$ and $\hat k_0=n-1$ in (\ref{shifting}) with the upper sign to obtain the identity
\begin{equation}
	\prod_{\hat j=0}^{m-1}\prod_{\hat k=0}^{n-1}\fft{\theta_1\left(\Delta_a-2\pi\fft{\hat j+\hat k\tilde\tau}m;\tau\right)}{\theta_1\left(\Delta_a+2\pi\fft{\hat j+\hat k\tilde\tau}m;\tau\right)}=e^{i[(N+1)\pi+(n-1)m\Delta_a]}.
\label{Ratio:pm}
\end{equation}
Inserting this into (\ref{After:shifting}), taking the product over $a$ and reducing the exponent then gives $v=(N+1)\pi$, which is also compatible with the solution (\ref{First:solution}) of \cite{Hosseini:2016cyf}.

As a result, we have found multiple solutions to the BAEs, (\ref{eq:full BAE}), labeled by three integers $m$, $n$, and $r$ such that $N=mn$ and $r=0,\ldots,n-1$. While we have not proven that these are the complete set of solutions to the BAEs (up to permutations), we argue below in section~\ref{sec:3} that they are in fact complete based on modular covariance of the index.

\subsection{The topologically twisted index}

We now compute the topologically twisted index for a particular sector labeled by $\{m,n,r\}$ by inserting the solution (\ref{eq:full BAE}) into (\ref{generating functional}).  Making this substitution gives
\begin{align}
	Z_{\{m,n,r\}}&=\fft{\mathcal A}{\det\mathbb B_{\{m,n,r\}}}\prod_{a=1}^3\left[\prod_{\hat j_1\hat k_1\ne \hat j_2\hat k_2}^N\frac{\theta_1\left(2\pi\frac{(\hat j_1-\hat j_2)+(\hat k_1-\hat k_2)\tilde\tau}m;\tau\right)}{\theta_1\left(\Delta_a+2\pi\frac{(\hat j_1-\hat j_2)+(\hat k_1-\hat k_2)\tilde\tau}m;\tau\right)}\right]^{1-\mathfrak n_a}\nn\\
	&=\fft{\mathcal A}{\det\mathbb B_{\{m,n,r\}}}\prod_{a=1}^3\left[\prod_{\hat j_2=0}^{m-1}\,\prod_{\hat k_2=0}^{n-1}\,\sideset{}{'}\prod_{\hat j_1=-\hat j_2}^{m-\hat j_2-1}\,\sideset{}{'}\prod_{\hat k_1=-\hat k_2}^{n-\hat k_2-1}\frac{\theta_1\left(2\pi\frac{\hat j_1+\hat k_1\tilde\tau}m;\tau\right)}{\theta_1\left(\Delta_a+2\pi\frac{\hat j_1+\hat k_1\tilde\tau}m;\tau\right)}\right]^{1-\mathfrak n_a},
\end{align}
where the primes indicate that $\hat j_1=\hat k_1=0$ is to be omitted from the double product. The product over $\hat j_1$ and $\hat k_1$ can be shifted using (\ref{shifting}) as follows:
\begin{equation}
	\,\sideset{}{'}\prod_{\hat j_1=-\hat j_2}^{m-\hat j_2-1}\,\sideset{}{'}\prod_{\hat k_1=-\hat k_2}^{n-\hat k_2-1}\frac{\theta_1\left(2\pi\frac{\hat j_1+\hat k_1\tilde\tau}m;\tau\right)}{\theta_1\left(\Delta_a+2\pi\frac{\hat j_1+\hat k_1\tilde\tau}m;\tau\right)}=e^{-im\hat k_2\Delta_a}\,\sideset{}{'}\prod_{\hat j_1=0}^{m-1}\,\sideset{}{'}\prod_{\hat k_1=0}^{n-1}\frac{\theta_1\left(2\pi\frac{\hat j_1+\hat k_1\tilde\tau}m;\tau\right)}{\theta_1\left(\Delta_a+2\pi\frac{\hat j_1+\hat k_1\tilde\tau}m;\tau\right)}.
\end{equation}
As a result, we have
\begin{equation}
	Z_{\{m,n,r\}}=\fft{\mathcal A}{\det\mathbb B_{\{m,n,r,s\}}}\prod_{a=1}^3\left[e^{-i\fft{m(n-1)}2\Delta_a}\sideset{}{'}\prod_{\hat j_1=0}^{m-1}\,\sideset{}{'}\prod_{\hat k_1=0}^{n-1}\frac{\theta_1\left(2\pi\frac{\hat j_1+\hat k_1\tilde\tau}m;\tau\right)}{\theta_1\left(\Delta_a+2\pi\frac{\hat j_1+\hat k_1\tilde\tau}m;\tau\right)}\right]^{N(1-\mathfrak n_a)}.
\label{eq:Zsimp}
\end{equation}
The product of the theta functions can be simplified by using the product form of $\theta_1(u;\tau)$ given in (\ref{eq:theta1}). We find
\begin{equation}
	\sideset{}{'}\prod_{\hat j=0}^{m-1}\,\sideset{}{'}\prod_{\hat k=0}^{n-1}\frac{\theta_1\left(u+2\pi\frac{\hat j+\hat k\tilde\tau}m;\tau\right)}{\eta(\tau)}=e^{i\fft{n-1}2\pi}e^{-\fft{i\pi nr}{6}}e^{-i\fft{m(n-1)}2u}\tilde q^{-\fft{(n-1)(n-1/2)}6}\fft{\eta(\tau)}{\theta_1(u;\tau)}\fft{\theta_1(mu;\tilde\tau)}{\eta(\tilde\tau)},
\label{eq:thetaprod}
\end{equation}
where $\tilde q=e^{2\pi i\tilde\tau}$. Moreover, taking the limit $u\to0$ and using $\theta_1'(0;\tau)=\eta(\tau)^3$ gives
\begin{equation}
   \sideset{}{'}\prod_{\hat j=0}^{m-1}\,\sideset{}{'}\prod_{\hat k=0}^{n-1}\frac{\theta_1(2\pi\frac{\hat j+\hat k\tilde\tau}m;\tau)}{\eta(\tau)}=e^{i\fft{n-1}2\pi}e^{-\fft{i\pi nr}{6}}\tilde q^{-\fft{(n-1)(n-1/2)}6}\fft{m\eta(\tilde\tau)^2}{\eta(\tau)^2}.
\end{equation}
Substituting these expressions and (\ref{eq:Afactor}) into (\ref{eq:Zsimp}) then gives
\begin{equation}
    Z_{\{m,n,r\}}=\fft{i^{N-1}}{\det\mathbb B_{\{m,n,r\}}}\prod_{a=1}^3\left[\left(\fft{\theta_1(\Delta_a;\tau)}{\eta(\tau)^3}\right)\left(\fft{m\eta(\tilde\tau)^3}{\theta_1(m\Delta_a;\tilde\tau)}\right)^N\right]^{1-\mathfrak n_a}.
\label{Index:exact:noBtilde}
\end{equation}

We now turn to the Jacobian matrix $\mathbb B_{\{m,n,r\}}$ given in (\ref{B matrix}). For the moment, we find it is convenient to maintain the original single index notation for the $u_j$'s.  Noting that (\ref{B matrix}) singles out $u_0$ as the constrained variable, the entries of the matrix are
\begin{subequations}
\begin{align}
	\mathbb B_{\mu,\nu}&\equiv\frac{\partial B_\mu}{\partial u_\nu}=\delta_{\mu\nu}\bigg(\sum_{j=0}^{N-1}g(u_\mu-u_j;\Delta_a,\tau)\bigg)-g(u_\mu-u_\nu;\Delta_a,\tau)+g(u_\mu-u_0;\Delta_a,\tau),\label{N-1 B matrix}\\
	\mathbb B_{0,\nu}&\equiv\frac{\partial B_0}{\partial u_\nu}=-\bigg(\sum_{j=0}^{N-1}g(u_0-u_j;\Delta_a,\tau)\bigg)-g(u_0-u_\nu;\Delta_a,\tau)+g(0;\Delta_a,\tau),\\
	\mathbb B_{\mu,0}&\equiv\frac{\partial B_\mu}{\partial v}=1,\\
	\mathbb B_{0,0}&\equiv\frac{\partial B_0}{\partial v}=1,
\end{align}\label{N B matrix}
\end{subequations}
where $\mu,\nu\in\{1,2,\ldots,N-1\}$. Here we have defined
\begin{equation}
	g(u;\Delta_a,\tau)\equiv i\sum_{a=1}^3\fft\partial{\partial\Delta_a}\log\bigl[\theta_1(\Delta_a+u;\tau)\theta_1(\Delta_a-u;\tau)\bigr].
\end{equation}
Since $g(u;\Delta_a,\tau)$ is an even function of $u$, we can derive the identities
\begin{equation}
	\sum_{j=0}^{N-1}\mathbb B_{j,\nu}=0\qquad\mbox{and}\qquad\sum_{j=0}^{N-1}\mathbb B_{j,0}=N.
\end{equation}
Consequently, we have
\begin{equation}
	\det\mathbb B=N\det\bigg[\frac{\partial(B_1,\cdots,B_{N-1})}{\partial(u_1,\cdots,u_{N-1})}\bigg].
	\label{B matrix:N-1}
\end{equation}
Therefore it is enough to study the determinant of the $(N-1)\times(N-1)$-square matrix whose entries are given by (\ref{N-1 B matrix}).

At this stage, we return to index pair notation given in (\ref{eq:full BAE}) by 
\begin{equation}
    u_{n\hat j+\hat k}\quad\to\quad u_{\hat j\hat k}=\bar u+2\pi\fft{\hat j+\hat k\tilde\tau}m,
    \label{Index:pair}
\end{equation}
which maps $\{u_j:j=0,\cdots,N-1\}$ onto $\{u_{\hat j\hat k}:\hat j=0,\cdots,m-1,~\hat k=0,\cdots,n-1\}$. We then define the $\mathcal G$-function as
\begin{equation}
	\mathcal G_{\{m,n,r\}}(\hat j,\hat k;\Delta_a,\tau)\equiv i\sum_{a=1}^3\fft\partial{\partial\Delta_a}\log\left[\theta_1\Bigl(\Delta_a+2\pi\fft{\hat j+\hat k\tilde\tau}m;\tau\Bigr)\theta_1\Bigl(\Delta_a-2\pi\fft{\hat j+\hat k\tilde\tau}m;\tau\Bigr)\right],
	\label{eq:Gdef}
\end{equation}
which yields
\begin{equation}
	\mathcal G_{\{m,n,r\}}(\hat j-\hat j_0,\hat k-\hat k_0;\Delta_a,\tau)=g(u_{\hat j\hat k}-u_{\hat j_0\hat k_0};\Delta_a,\tau).
\end{equation}
Accordingly, the sum in (\ref{N-1 B matrix}) can be written in terms of index pair notation as
\begin{equation}
    \sum_{j=0}^{N-1}g(u_j-u_\mu;\Delta_a,\tau)\quad\to\quad\sum_{\hat j=0}^{m-1}\sum_{\hat k=0}^{n-1}\mathcal G_{\{m,n,r\}}(\hat j-\hat j_\mu,\hat k-\hat k_\mu;\Delta_a,\tau)\label{diagonal}
\end{equation}
where $\mu=n\hat j_\mu+\hat k_\mu$. Now changing the summation over $\hat j$ and $\hat k$ into a product within the log and inserting (\ref{shifting}) then gives 
\begin{align}
	\sum_{\hat j=0}^{m-1}\sum_{\hat k=0}^{n-1}\mathcal G_{\{m,n,r\}}(\hat j-\hat j_\mu,\hat k-\hat k_\mu;\Delta_a,\tau)=\sum_{\hat j=0}^{m-1}\sum_{\hat k=0}^{n-1}\mathcal G_{\{m,n,r\}}(\hat j,\hat k;\Delta_a,\tau)
\end{align}
so that the sum in (\ref{N-1 B matrix}) is in fact independent of which entry $\mu$ is being considered. Simplifying the product of theta functions within the log using (\ref{eq:thetaprod}), we get
\begin{align}\label{sum:G}
	\sum_{\hat j=0}^{m-1}\sum_{\hat k=0}^{n-1}\mathcal G_{\{m,n,r\}}(\hat j,\hat k;\Delta_a,\tau)=2i\sum_{a=1}^3\partial_{\Delta_a}\log\theta_1(m\Delta_a;\tilde\tau)
\end{align}
where the prime denotes differentiation with respect to the first argument of $\theta_1$. Finally, inserting (\ref{sum:G}) into (\ref{N-1 B matrix}), we can rewrite (\ref{N-1 B matrix}) as
\begin{equation}
    [\mathbb B_{\{m,n,r\}}]_{\mu,\nu}=\left(2i\sum_{a=1}^3\partial_{\Delta_a}\log\theta_1\left(m\Delta_a;\tilde\tau\right)\right)\left[I_{N-1}+\tilde{\mathbb B}_{\{m,n,r\}}\right]_{\mu,\nu}
\label{eq:BBmunu}
\end{equation}
where $\tilde{\mathbb B}_{\{m,n,r\}}$ is an $(N-1)\times(N-1)$ square matrix with entries
\begin{equation}\label{B:tilde:entry}
	[\tilde{\mathbb B}_{\{m,n,r\}}]_{\mu,\nu}=\fft{\mathcal G_{\{m,n,r\}}(\hat j_\mu,\hat k_\mu;\Delta_a,\tau)-\mathcal G_{\{m,n,r\}}(\hat j_\mu-\hat j_\nu,\hat k_\mu-\hat k_\nu;\Delta_a,\tau)}{2i\sum_{a=1}^3\partial_{\Delta_a}\log\theta_1\left(m\Delta_a;\tilde\tau\right)}.
\end{equation}
Then (\ref{B matrix:N-1}) leads to
\begin{equation}
    \det\mathbb B_{\{m,n,r\}}=N\left(2i\sum_{a=1}^3\partial_{\Delta_a}\log\theta_1\left(m\Delta_a;\tilde\tau\right)\right)^{N-1}\det(1+\tilde{\mathbb B}_{\{m,n,r\}}).
\label{eq:detBexp}
\end{equation}

Finally, the contribution to the topologically twisted index from the sector labeled by $\{m,n,r\}$ is given by combining (\ref{Index:exact:noBtilde}) with (\ref{eq:detBexp}),
\begin{equation}\label{Index:exact}
	Z_{\{m,n,r\}}(\tau;\Delta_a,\mathfrak n_a)=\frac{\displaystyle \prod_{a=1}^3\left[\left(\fft{\theta_1(\Delta_a;\tau)}{\eta(\tau)^3}\right)\left(\fft{m\eta(\tilde\tau)^3}{\theta_1\left(m\Delta_a;\tilde\tau\right)}\right)^N\right]^{1-\mathfrak n_a}}{\displaystyle N\det(1+\tilde{\mathbb B}_{\{m,n,r\}})\left[2\sum_{a=1}^3\partial_{\Delta_a}\log\theta_1\left(m\Delta_a;\tilde\tau\right)\right]^{N-1}}.
\end{equation}
%

\section{The index as an elliptic genus}
\label{sec:3}

As we have seen above, there are multiple solutions to the BAEs, each labeled by a set of integers $\{m,n,r\}$, corresponding to the modding out of the original $T^2$ by a freely acting $\mathbb Z_m\times\mathbb Z_n$ action.  The sum over these multiple solutions $I\in\mathrm{BAE}s$ in (\ref{generating functional}) is non-trivial, and explicitly takes the form
\begin{equation}
	Z(\tau;\Delta_a,\mathfrak n_a)=\sum_{n|N}\sum_{r=0}^{n-1} Z_{\{N/n,n,r\}}(\tau;\Delta_a,\mathfrak n_a),
	\label{eq:Ztotsum}
\end{equation}
where $Z_{\{m,n,r\}}$ is given in (\ref{Index:exact}). In this section, we study this expression further for arbitrary $\tau$ and $N$. In particular, we show explicitly that the index is an elliptic genus, which can be seen based on reduction over the $S^2$ \cite{Honda:2015yha}. Here the sum in (\ref{eq:Ztotsum}) is crucial to ensure proper modular behavior of the index, since modular transformations permute the individual sectors labeled by $\{m,n,r\}$.

For example, consider the case $N=6$, where the index (\ref{eq:Ztotsum}) is a sum over the twelve sectors
\begin{align}
	\{m,n,r\}=&\{1,6,0\},\{1,6,1\},\{1,6,2\},\{1,6,3\},\{1,6,4\},\{1,6,5\},\nn\\&\{2,3,0\},\{2,3,1\},\{2,3,2\},\{3,2,0\},\{3,2,1\},\{6,1,0\},
\label{N=6:sectors}
\end{align}
with corresponding modular parameters
\begin{align}\label{modded:tau}
	\tilde\tau=\fft\tau6,\fft{\tau+1}6,\fft{\tau+2}6,\fft{\tau+3}6,\fft{\tau+4}6,\fft{\tau+5}6,\fft{2\tau}3,\fft{2\tau+1}3,\fft{2\tau+2}3,\fft{3\tau}2,\fft{3\tau+1}2,6\tau.
\end{align}
These modular parameters are closed under $T$: $\tau\to\tau+1$ according to
\begin{align}
	\left(\fft\tau6,\fft{\tau+1}6,\fft{\tau+2}6,\fft{\tau+3}6,\fft{\tau+4}6,\fft{\tau+5}6\right)\left(\fft{2\tau}3,\fft{2\tau+2}3,\fft{2\tau+1}3\right)\left(\fft{3\tau}2,\fft{3\tau+1}2\right)\left(6\tau\right),
	\label{eq:6T}
\end{align}
and under $S$: $\tau\to-1/\tau$ according to
\begin{align}
	\left(\fft\tau6,6\tau\right)\left(\fft{\tau+1}6,\fft{\tau+5}6\right)\left(\fft{\tau+2}6,\fft{2\tau+2}3\right)\left(\fft{\tau+3}6,\fft{3\tau+1}2\right)\left(\fft{\tau+4}6,\fft{2\tau+1}3\right)\left(\fft{2\tau}3,\fft{3\tau}2\right).
\end{align}
Obtaining the orbit under $T$ is straightforward, while obtaining that under $S$ is somewhat more involved.  Consider, for example, the action of $S$ on the $\{1,6,2\}$ sector, with $\tilde\tau=(\tau+2)/6$.  We first take $S$: $\tilde\tau\to\tilde\tau'=(2\tau-1)/6\tau$, and then perform a $SL(2;\mathbb Z)$ transformation $\tilde\tau'\to(2\tilde\tau'-1)/(3\tilde\tau'-1)$ to bring this into the form $(2\tau+2)/3$, corresponding to the $\{2,3,2\}$ sector.  Of course, the detailed modular properties of the topologically twisted index depends on how precisely the various building blocks of $Z_{\{m,n,r\}}$ transform.

Before considering the general case, we gain additional insight from the example of $N=2$.  In this case, there are only three sectors, denoted by $\{1,2,0\}$, $\{1,2,1\}$ and $\{2,1,0\}$.  The topologically twisted index is then given by the sum
\begin{equation}
	Z^{N=2}(\tau;\Delta_a,\mathfrak n_a)=\sum_{i=2}^4\left(\frac1{8\sum_a\partial_{\Delta_a}\log\theta_i(\Delta_a;\tau)}\prod_a\left[\fft{\eta(\tau)^3}{\theta_1(\Delta_a;\tau)}\left(\fft{\theta_i(0;\tau)}{\theta_i(\Delta_a;\tau)}\right)^2\right]^{1-n_a}\right),
\label{eq:ZN2}
\end{equation}
where $i=2,3,4$ correspond to the $\{2,1,0\}$, $\{1,2,1\}$ and $\{1,2,0\}$ sectors, respectively. Then the modular properties of the index can be derived from those of the elliptic theta functions, $\theta_i$.

Turning to the general case, for the index to be an elliptic genus, it must transform as a weak Jacobi form of weight zero.  Here it is worth recalling that, for a single chemical potential, a Jacobi form of weight $k$ and index $m$ transforms according to
\begin{subequations}
	\begin{align}
	\phi(\tau,u+2\pi(\lambda\tau+\mu))&=(-1)^{2m(\lambda+\mu)}q^{-m\lambda^2}e^{-2im\lambda u}\phi(\tau,u),
	\label{eq:jacobi1}\\
	\phi\left(\fft{a\tau+b}{c\tau+d},\fft{u}{c\tau+d}\right)&=(c\tau+d)^ke^{\fft{imcu^2}{2\pi(c\tau+d)}}\phi(\tau,u).
	\label{eq:jacobi2}
	\end{align}
\end{subequations}
It is straightforward to generalize this to the case of three chemical potentials, and we verify below that the index (\ref{eq:Ztotsum}) indeed transforms as a weak Jacobi form of weight zero and indices
\begin{equation}
	m_a=-\fft{N^2-1}2(1-\mathfrak n_a),
\label{eq:indices}
\end{equation}
under the constraint $\sum_a\Delta_a=0$. To do so, we first consider the periodic shifts $\Delta_a\to\Delta_a+2\pi$ and $\Delta_a\to\Delta_a+2\pi\tau$ for (\ref{eq:jacobi1}), and next consider the modular transformations $T:\tau\to\tau+1$ and $S:\tau\to-1/\tau$ for (\ref{eq:jacobi2}).  Note that the index $m_a$ is a half-integer when both $N$ and $\mathfrak n_a$ are even, and an integer otherwise.

\subsection{Periodic shifts of $\Delta_a$}

We first consider the shift $\Delta_{\hat a}\to\Delta_{\hat a}+2\pi$ for a single $\Delta_{\hat a}$.  Since $\theta_1$ picks up a minus sign for every $2\pi$ shift, the numerator of (\ref{Index:exact:noBtilde}) picks up a sign $(-1)^{(1-mN)(1-\mathfrak n_{\hat a})}$, while the denominator is unchanged since the logarithmic derivatives of $\theta_1$ are not sensitive to the sign.
As a result, we find
\begin{equation}
    Z_{\{m,n,r\}}\to(-1)^{(1-mN)(1-\mathfrak n_{\hat a})}Z_{\{m,n,r\}}=(-1)^{2m_{\hat a}}(-1)^{N(N-m)(1-\mathfrak n_{\hat a})}Z_{\{m,n,r\}},
\end{equation}
where we substituted in the index $m_{\hat a}$ from (\ref{eq:indices}).  Writing $N=mn$ then gives $N(N-m)=m^2n(n-1)$, which is an even integer.  Thus the second factor above is simply $+1$, and we are left with $Z_{\{m,n,r\}}\to(-1)^{2m_{\hat a}}Z_{\{m,n,r\}}$, in agreement with (\ref{eq:jacobi1}).  Note that this result is valid even if we only shift a single $\Delta_{\hat a}$.

For the shift $\Delta_{\hat a}\to\Delta_{\hat a}+2\pi\tau$, we first consider the numerator factors in (\ref{Index:exact}) using (\ref{eq:uperiod}).  For $\theta_1(\Delta_{\hat a},\tau)$, we find simply
\begin{equation}
    \theta_1(\Delta_{\hat a}+2\pi\tau,\tau)=-q^{-1/2}y_{\hat a}^{-1}\theta_1(\Delta_{\hat a},\tau),
    \label{eq:t1t}
\end{equation}
where $y_{\hat a}=e^{i\Delta_{\hat a}}$.  For $\theta_1(m\Delta_{\hat a},\tilde\tau)$, we first write
\begin{equation}
    \theta_1(m(\Delta_{\hat a}+2\pi\tau),\tilde\tau)=\theta_1(m\Delta_{\hat a}+2\pi(n\tilde\tau-r),\tilde\tau)=(-1)^{r+n}\tilde q^{-n^2/2}y_{\hat a}^{-N}\theta_1(m\Delta_{\hat a},\tilde\tau),
\end{equation}
and use the relation $\tilde q^n=e^{2\pi in\tilde\tau}=e^{2\pi ir}q^m$ to obtain
\begin{equation}
    \theta_1(m(\Delta_{\hat a}+2\pi\tau),\tilde\tau)=(-1)^{n+r(n+1)}q^{-N/2}y_{\hat a}^{-N}\theta_1(m\Delta_{\hat a},\tilde\tau).
    \label{eq:t1mt}
\end{equation}
This demonstrates that the numerator picks up an overall factor
\begin{equation}
    \left[(-1)^{1-N(n+r(n+1))}q^{(N^2-1)/2}y_{\hat a}^{N^2-1}\right]^{1-\mathfrak n_{\hat a}},
    \label{eq:numershift}
\end{equation}
under a shift of $\Delta_{\hat a}$ by $2\pi\tau$.  As above, the sign factor can be rewritten as
\begin{align}
    1-N(n+r(n+1))&=-(N^2-1)+N(n(m-1)-r(n+1))\nn\\
    &=-(N^2-1)+n^2m(m-1)-rmn(n+1).
\end{align}
Since the last two terms in the final expression are even, they do not contribute to the overall sign, and we are left with
\begin{equation}
    Z_{\{m,n,r\}}^{\mathrm{numer}}\to(-1)^{2m_{\hat a}}q^{-m_{\hat a}}y_{\hat a}^{-2m_{\hat a}}Z_{\{m,n,r\}}^{\mathrm{numer}},
\end{equation}
which is the expected result for a Jacobi form of index 
$m_{\hat a}$ given by (\ref{eq:indices}).

Since the numerator by itself transforms properly under the shift of $\Delta_{\hat a}$ by $2\pi\tau$, we see that the denominator must be inert under this shift.  This is not entirely obvious, though, as the logarithmic derivatives of $\theta_1$ transform as
\begin{align}
    \partial_{\Delta_{\hat a}}\log\theta_1(\Delta_{\hat a}+2\pi\tau,\tau)&=\partial_{\Delta_{\hat a}}\log\theta_1(\Delta_{\hat a},\tau)-i,\nn\\
    \partial_{\Delta_{\hat a}}\log\theta_1(m(\Delta_{\hat a}+2\pi\tau),\tilde\tau)&=\partial_{\Delta_{\hat a}}\log\theta_1(m\Delta_{\hat a},\tilde\tau)-iN,
\end{align}
as can be seen directly from (\ref{eq:t1t}) and (\ref{eq:t1mt}).  The sum of logarithmic derivatives, however, is invariant so long as we simultaneously shift another chemical potential, say $\Delta_{\hat b}$, by $-2\pi\tau$, since then these additional factors will cancel. Therefore the denominator is invariant under this combined shift, and hence (\ref{eq:numershift}) extends to $Z_{\{m,n,r\}}$ itself.  Note that this simultaneous shift is in fact required to maintain the condition that the $\Delta_{\hat a}$'s sum to $2\pi\mathbb Z$.

\subsection{Modular transformations}

We now turn to the properties of the topologically twisted index under modular transformations.  Since a general transformation can be generated by a combination of $T$ and $S$, it is sufficient for us to demonstrate the following properties:
\begin{subequations}
	\begin{align}
	T:&\quad Z(\tau+1;\Delta_a,\mathfrak n_a)=Z(\tau;\Delta_a,\mathfrak n_a),\label{T:inv}\\
	S:&\quad Z(-1/\tau;\Delta_a/\tau,\mathfrak n_a)=e^{\fft{i}{2\pi\tau}\sum_{a=1}^3m_a\Delta_a^2}Z(\tau;\Delta_a,\mathfrak n_a).\label{S:inv}
	\end{align}
\end{subequations}
These follow from the definition (\ref{eq:jacobi2}) for a Jacobi form of weight zero and indices $m_a$ for the chemical potentials $\Delta_a$.

\subsubsection{$T$ transformation}

We begin with the $T$ transformation.  As indicated in (\ref{T:inv}), we expect the partition function to be invariant under $T$.  Nevertheless, the individual sectors labeled by $\{m,n,r\}$ will get permuted, as in the $N=6$ example shown in (\ref{eq:6T}).  We thus work one sector at a time, and in particular consider the $T$ transformation of $Z_{\{m,n,r\}}$.

To proceed, we consider the expression (\ref{Index:exact:noBtilde}), and observe that the numerator is built from the combination
\begin{equation}
    \psi(u;\tau)\equiv\fft{\theta_1(u;\tau)}{\eta(\tau)^3},
    \label{eq:psiut}
\end{equation}
which transforms as a weak Jacobi form of weight $-1$ and index $1/2$, as can be seen from (\ref{eq:sl2z}).  For $\psi(\Delta_a;\tau)$, we have simply
\begin{equation}
    T:\psi(\Delta_a;\tau)\to\psi(\Delta_a;\tau).
    \label{eq:Ttau}
\end{equation}
However, the transformation is not as direct for $\psi(m\Delta_a;\tilde\tau)$, since $T:\tilde\tau\to\tilde\tau+m/n$, which is not a $SL(2,\mathbb Z)$ transformation on $\tilde\tau$.  In this case, it is more useful to note that
\begin{equation}
    T:\fft{m\tau+r}n\to\fft{m\tau+(r+m)}n=\fft{m\tau+r'}n+\left\lfloor\fft{r+m}n\right\rfloor,
\end{equation}
where $r'=r+m\pmod n$.  Since $\psi$ is invariant under integer shifts of the modular parameter, we end up with
\begin{equation}
    T:\psi(m\Delta_a;\tilde\tau)\to\psi(m'\Delta_a;\tilde\tau'),
    \label{eq:Ttildetau}
\end{equation}
where
\begin{equation}
    \tilde\tau'\equiv\fft{m'\tau+r'}{n'},\qquad \{m',n',r'\}=\{m,n,r+m\kern-.5em\pmod n\}.
\end{equation}
The combination of (\ref{eq:Ttau}) and (\ref{eq:Ttildetau}) then demonstrates the simple transformation
\begin{equation}
    T:Z_{\{m,n,r\}}^{\mathrm{numer}}\to Z_{\{m',n',r'\}}^{\mathrm{numer}},
\end{equation}
as anticipated in (\ref{eq:6T}).

To be complete, we must also investigate the $T$ transformation on the denominator of (\ref{Index:exact:noBtilde}), which comes from the determinant of $\mathbb B_{\{m,n,r\}}$.  Here we use the double periodicity (\ref{eq:uperiod}) and the modular property (\ref{modular of theta}), to obtain the map
\begin{equation}
	\mathcal G_{\{m,n,r\}}(\hat j,\hat k;\Delta_a,\tau+1)=\mathcal G_{\{m',n',r'\}}(\hat j',\hat k';\Delta_a,\tau),
	\label{G-fct:T}
\end{equation}
with
\begin{equation}
	\hat j'=\hat j+\hat k\left\lfloor\fft{r+m}{n}\right\rfloor\pmod m,\qquad\hat k'=\hat k.
	\label{T:inv:G}
\end{equation}
Then since the above $(\hat j,\hat k)\to(\hat j',\hat k')$ is a bijective map from $\mathbb Z_m\times\mathbb Z_n$ to $\mathbb Z_{m'}\times\mathbb Z_{n'}$, we get (see Appendix~\ref{App E})
\begin{equation}
	T:\det\mathbb B_{\{m,n,r\}}\to\det\mathbb B_{\{m',n',r'\}},
	\label{T:inv:condi2}
\end{equation}
and hence the denominator transforms in the expected manner as well.  As a result, $T$ permutes the sectors without any additional factors, $T:Z_{\{m,n,r\}}\to Z_{\{m',n',r'\}}$.  Finally, since $\{m,n,r\}\to\{m',n',r'\}$ is bijective, it is clear that the full partition function (\ref{eq:Ztotsum}) is indeed invariant under $T$ transformations, (\ref{T:inv}).

\subsubsection{$S$ transformation}
\label{sec:S:inv}

We now turn to the $S$ transformation, which takes $\Delta_a\to\Delta_a/\tau$ along with $\tau\to-1/\tau$.  Once again, we start with the numerator.  Since $\psi(u;\tau)$ defined in (\ref{eq:psiut}) is a weak Jacobi form of weight $-1$ and index $1/2$, we immediately have
\begin{equation}
    S:\psi(\Delta_a;\tau)\to\fft1\tau e^{\fft{i\Delta_a^2}{4\pi\tau}}\psi(\Delta_a,\tau).
    \label{eq:Spsi}
\end{equation}
For $\psi(m\Delta_a;\tilde\tau)$, it is important to realize that $S$ does not simply take $\tilde\tau$ to $-1/\tilde\tau$.  Instead, we want to map $\tilde\tau$ into a new $\tilde\tau'$, at least up to a $SL(2;\mathbb Z)$ transformation.  In particular, we demand
\begin{equation}
    S:\fft{m\tau+r}n\to\fft{r\tau-m}{n\tau}=\fft{a\tilde\tau'+b}{c\tilde\tau'+d},
    \label{eq:Sabcd}
\end{equation}
where $\tilde\tau'=(m'\tau+r')/n'$.  The resulting $SL(2;\mathbb Z)$ transformation is given by
\begin{equation}
    a=\fft{r}g,\qquad c=\fft{n}g,\qquad ad-bc=1,\qquad g\equiv\gcd(n,r),
    \label{eq:abcd}
\end{equation}
and $\tilde\tau'$ takes the form
\begin{equation}
    \tilde\tau'=\fft{m'\tau+r'}{n'},\qquad\{m',n',r'\}=\{g,N/g,-dm\}.
    \label{eq:mpnprp}
\end{equation}
Here $b$ and $d$ are uniquely determined as the solution to (\ref{eq:abcd}) under the constraint for $r'$, $0\leq r'<n'$.  Also note that we can make use of the simple relation $c\tilde\tau'+d=m'\tau/m$, which can be derived without explicit knowledge of $b$ and $d$.  Given (\ref{eq:Sabcd}), we then find
\begin{equation}
    S:\psi(m\Delta_a;\tilde\tau)\to\psi\left(\fft{m\Delta_a}\tau;\fft{r\tau-m}{n\tau}\right)=\psi\left(\fft{m'\Delta_a}{c\tilde\tau'+d};\fft{a\tilde\tau'+b}{c\tilde\tau'+d}\right)=\fft{m}{m'\tau}e^{\fft{iN\Delta_a^2}{4\pi\tau}}\psi(m'\Delta_a,\tilde\tau').
\end{equation}
Inserting this expression along with (\ref{eq:Spsi}) into (\ref{Index:exact:noBtilde}) then gives
\begin{equation}
    S:Z_{\{m,n,r\}}^{\mathrm{numer}}\to\tau^{N-1}e^{\fft{i}{2\pi\tau}\sum_am_a\Delta_a^2}Z_{\{m',n',r'\}}^{\mathrm{numer}},
\end{equation}
with $m_a$ given in (\ref{eq:indices}).  

The extra factor of $\tau^{N-1}$ is canceled by a similar factor arising from $\det\mathbb B$ in the denominator.  For this determinant, we use the double periodicity (\ref{eq:uperiod}) and the modular property (\ref{modular of theta}), along with the requirement $\sum_a\Delta_a=0$ to obtain the map
\begin{equation}
	\mathcal G_{\{m,n,r\}}(\hat j,\hat k;\Delta_a/\tau,-1/\tau)=\tau\mathcal G_{\{m',n',r'\}}(\hat j',\hat k';\Delta_a,\tau)
	\label{G-fct:S}
\end{equation}
with
\begin{subequations}
	\begin{align}
	\hat j'&=-\fft{g}{n}(\hat k+d\hat k')\mod g,\label{S:inv:G:a}
	\\\hat k'&=\fft{n}{g}\hat j+\fft{r}{g}\hat k\mod\fft{N}{g}.\label{S:inv:G:b}
	\end{align}%
\label{S:inv:G}%
\end{subequations}
In Appendix~\ref{App D}, we show that the above $(\hat j,\hat k)\to(\hat j',\hat k')$ is a bijective map from $\mathbb Z_m\times\mathbb Z_n$ to $\mathbb Z_{m'}\times\mathbb Z_{n'}$.  Therefore, we get (see Appendix~\ref{App E})
\begin{equation}
	S:\det\mathbb B_{\{m,n,r\}}\to\tau^{N-1}\det\mathbb B_{\{m',n',r'\}},
	\label{S:inv:condi2}
\end{equation}
which cancels the extra factor of $\tau^{N-1}$ in the numerator. As a result, $S$ permutes the sectors with a common factor, $S:Z_{\{m,n,r\}}\to e^{\fft{i}{2\pi\tau}\sum_am_a\Delta_a^2}Z_{\{m',n',r'\}}$. Then since $\{m,n,r\}\to\{m',n',r'\}$ is self-inverse and therefore bijective, the full partition function (\ref{eq:Ztotsum}) transforms under $S$ transformation as (\ref{S:inv}).

Finally, we wish to explain why the chemical potentials must sum to zero in order for the index to be a proper modular form, in particular under the $S$-transformation: since $S$ takes $\Delta_a$ to $\Delta_a/\tau$, we must demand the simultaneous conditions
\begin{equation}
	\sum_{a=1}^3\Delta_a=2\pi\mathbb Z\quad\mbox{and}\quad\sum_{a=1}^3\Delta_a=2\pi\tau\mathbb Z
\end{equation}
to satisfy the first constraint given in (\ref{constraints}) for both $Z(\tau;\Delta_a,\mathfrak n_a)$ and $Z(-1/\tau;\Delta_a/\tau,\mathfrak n_a)$, which only makes sense when $\sum_{a=1}^3\Delta_a=0$. Of course, we can always use the second type of the degrees of freedom introduced in (\ref{Deg:free}), $\Delta_a\to\Delta_a+2\pi\mathbb Z$, to set $\sum_{a=1}^3\Delta_a=0$, so this is not a serious restriction on the index.

\section{The topologically twisted index in the `high-temperature' limit}
\label{sec:4}

Given the construction of the index as a sum over sectors, (\ref{eq:Ztotsum}), we now revisit the `high-temperature' limit, $\beta\to0^+$ with $\tau=i\beta/2\pi$, first investigated in \cite{Hosseini:2016cyf} for the single sector $Z_{\{1,N,0\}}$.  Note that, in what follows, we restrict to purely imaginary $\tau$, corresponding to a square torus, and real chemical potentials $\Delta_a$. In order to explore this limit, it is natural to perform an $S$ transformation (\ref{S:inv}) assuming $\sum_a\Delta_a=0$ so that the transformed modular parameter has large imaginary part.  In particular, we write
\begin{equation}
    Z(\tau;\Delta_a,\mathfrak n_a)=e^{\fft{i}{2\pi\tau'}\sum_am_a\Delta_a'^2}Z(\tau';\Delta_a',\mathfrak n_a),
    \label{eq:Zttp}
\end{equation}
where
\begin{equation}
    \tau'=-\fft1\tau=\fft{2\pi i}\beta,\qquad\Delta_a'=\fft{\Delta_a}\tau=-\fft{2\pi i\Delta_a}\beta.
\end{equation}
The partition function $Z(\tau';\Delta_a',\mathfrak n_a)$ receives contributions from individual sectors $Z_{\{m',n',r'\}}$ as we have seen in (\ref{eq:Ztotsum}), and we generically expect only one or a handful of sectors to dominate.  To see this, we first work on the expression for a fixed sector, and then look for the dominant contribution to the sum over sectors.

\subsection{Expanding $Z_{\{m',n',r'\}}$ in the `high-temperature' limit}

In order to expand $Z_{\{m',n',r'\}}$, we rewrite (\ref{Index:exact}) as
\begin{equation}
    Z_{\{m',n',r'\}}(\tau';\Delta_a',\mathfrak n_a)=\fft{\prod_a\left[\psi(\Delta_a';\tau')\psi(m'\Delta_a';\tilde\tau')^{-N}\right]^{1-\mathfrak n_a}}{n'\det\left(1+\tilde{\mathbb B}_{\{m',n',r'\}}\right)\left[2\sum_a\fft{\psi'(m'\Delta_a';\tilde\tau')}{\psi(m'\Delta_a';\tilde\tau')}\right]^{N-1}}.
    \label{Index:m'n'r'}
\end{equation}
The numerator can be easily treated using the asymptotic expression for $\psi$, (\ref{psi:highT}), as
\begin{subequations}\label{Numerator}
\begin{align}
	\psi(\Delta_a';\tau')&=-i(-1)^{D_a}\exp\left[\fft{i\Delta_a^2}{4\pi\tau}+\fft{i\pi}{\tau}d_a(1-d_a)\right]\left(1+\mathcal O(e^{-\fft{2\pi}{\tau_2}\min(d_a,1-d_a)})\right),\label{eq:term1}\\
	\psi(m'\Delta_a';\tilde\tau')&=-i(-1)^{X_a}e^{i\pi\fft{r'}{n'}X_a(X_a+1)}\exp\left[\fft{iN\Delta_a^2}{4\pi\tau}+\fft{i\pi m'}{n'\tau}x_a(1-x_a)\right]\nn\\
	&\kern14em\times\left(1+\mathcal O(e^{-\fft{2\pi}{\tau_2}\fft{m'}{n'}\min(x_a,1-x_a)})\right),\label{eq:term2}
\end{align}
\end{subequations}
where
\begin{subequations}
	\begin{align}
	&d_a\equiv\fft{\Delta_a}{2\pi}\pmod1,&D_a&\equiv\left\lfloor\fft{\Delta_a}{2\pi}\right\rfloor,\label{eq:d:def}\\
	&x_a\equiv\fft{n'\Delta_a}{2\pi }\pmod1,&X_a&\equiv\left\lfloor\fft{n'\Delta_a}{2\pi }\right\rfloor.\label{eq:x:def}
	\end{align}
\end{subequations}
Note that these expressions break down if $d_a=0$ or $x_a=0$, so from now on we assume $d_a$'s are not integer multiples of $1/n'$ where this does not occur.

For the denominator, we first examine the logarithmic derivative term in (\ref{Index:m'n'r'}). So long as we avoid the special cases $x_a=0$, the asymptotic expression (\ref{eq:term2}) is differentiable with respect to its first argument, and we obtain
\begin{equation}
	2\sum_a\fft{\psi'(m'\Delta_a';\tilde\tau')}{\psi(m'\Delta_a';\tilde\tau')}=i\sum_{a=1}^3(1+2X_a)+\mathcal O(e^{-\fft{2\pi}{\tau_2}\fft{m'}{n'}\min(x_a,1-x_a)}).
	\label{eq:term3}
\end{equation}
Since $\sum_a\Delta_a=0$ and we avoid special cases, we see that $X_a$ must generically sum to either $-1$ or $-2$. Therefore (\ref{eq:term3}) is in fact just $\pm i$.

The remaining term, namely $\det\relax(1+\tilde{\mathbb B}_{\{m',n',r'\}})$, is more difficult to analyze.  So for the moment we leave it implicit.  In this case, combining the numerator terms (\ref{Numerator}) with (\ref{eq:term3}) and taking into account the prefactor in (\ref{eq:Zttp}) gives
\begin{align}
	\log Z_{\{m,n,r\}}(\tau;\Delta_a,\mathfrak n_a)=\,&\fft{2\pi^2}{\beta}\sum_{a=1}^3(1-\mathfrak n_a)\left(d_a(1-d_a)-m'^2x_a(1-x_a)\right)-\log n'\nn\\
    &-\log\det(1+\tilde{\mathbb B}_{\{m',n',r'\}})+i\varphi\nn\\
    &+\mathcal O(e^{-\fft{4\pi^2}\beta\min(d_a,1-d_a)},e^{-\fft{4\pi^2}\beta\fft{m'}{n'}\min(x_a,1-x_a)}),
    \label{Index:highT:mnr}
\end{align}
where $\varphi$ is a phase independent of $\tau$, and the transformed quantities $\{m',n',r'\}$ are given by (\ref{eq:abcd}) and (\ref{eq:mpnprp}).

\subsection{Examination of the determinant factor}

The asymptotic expression for the index, (\ref{Index:highT:mnr}), is now complete up to the expansion of the determinant.  Unfortunately, its structure is rather intricate, and we have been unable to find a simple universal formula describing its asymptotics.  The main issue is the observation that the high temperature limit of $\log\det\relax(1+\tilde{\mathbb B}_{\{m',n',r'\}})$ can be of either $\mathcal O(1)$ or $\mathcal O(1/\beta)$.  This term is relatively unimportant in the former case, but will contribute to the leading order behavior in (\ref{Index:highT:mnr}) in the latter case.  However, which case the determinant factor is in depends in a non-obvious manner on the chemical potentials $\Delta_a$ and is not easily obtained.

We now proceed with a closer look at the matrix $\tilde{\mathbb B}_{\{m',n',r'\}}$ defined in (\ref{B:tilde:entry}).  To avoid unnecessary notation, we will omit the universal arguments $(\Delta_a',\tau')=(\Delta_a/\tau,-1/\tau)$ and occasionally the sector labels $\{m',n',r'\}$, in what follows.  In this case, the $\tilde{\mathbb B}$ matrix entries can be simply written as
\begin{equation}
	[\tilde{\mathbb B}_{\{m',n',r'\}}]_{\mu,\nu}=\fft{\mathcal G_{\{m',n',r'\}}(\hat j'_\mu,\hat k'_\mu)-\mathcal G_{\{m',n',r'\}}(\hat j'_\mu-\hat j'_\nu,\hat k'_\mu-\hat k'_\nu)}{\sum_{\hat j'=0}^{m'-1}\sum_{\hat k'=0}^{n'-1}\mathcal G_{\{m',n',r'\}}(\hat j',\hat k')},
	\label{Btilde:entry}
\end{equation}
where we have the index pair associations $\mu\to(\hat j'_\mu,\hat k'_\mu)$ and $\nu\to(\hat j'_\nu,\hat k'_\nu)$.  At this stage it is convenient to note that while this is originally an $(N-1)\times(N-1)$ square matrix, it can be extended to an $N\times N$ square matrix by including the $\mu=0$ and $\nu=0$ entries.  This is equivalent to allowing $\hat j'$ and $\hat k'$ to independently run over $0\ldots m'-1$ and $0\ldots n'-1$ without removing the $(0,0)$ pair.  Since the first column of $\tilde{\mathbb B}$ with entries $[\tilde{\mathbb B}]_{\mu,0}$ vanishes identically, however, the determinant $\det\relax(1+\tilde{\mathbb B})$ can be viewed either as an $(N-1)\times(N-1)$ or an $N\times N$ determinant.

Taking the logarithmic derivative of $\psi(u;\tau)$ and using the asymptotic expansion (\ref{psi:highT}) gives the high-temperature expansion of $\mathcal G$, which is necessary to study the high-temperature limit of the matrix $\tilde{\mathbb B}_{\{m',n',r'\}}$ with entries (\ref{Btilde:entry}).  We keep the $\mathcal O(1)$ and the first subexponential term, $\mathcal G(\hat j',\hat k')=\mathcal G^0(\hat j',\hat k')+\mathcal G^{\mathrm{exp}}(\hat j',\hat k')+\cdots$, where
\begin{subequations}
	\begin{align}
	\mathcal G^0(\hat j',\hat k')=&-\sum_{a=1}^3\left(1+D_a(\hat k'/n')+D_a(-\hat k'/n')\right),
	\label{G-fct:leading}
	\\\mathcal G^{\mathrm{exp}}(\hat j',\hat k')=&\sum_{a=1}^3\sum_{\sigma=\pm}\left(e^{-\fft{2\pi i\sigma}N(\hat j'n'+\hat k'r')}e^{-\fft{2\pi}{\tau_2}(1-d_a(\sigma\hat k'/n'))}-e^{\fft{2\pi i\sigma}N(\hat j'n'+\hat k'r')}e^{-\fft{2\pi}{\tau_2}d_a(\sigma\hat k'/n')}\right),
	\label{G-fct:exp}
	\end{align}
\end{subequations}\label{G-fct}
and we have defined
\begin{equation}
	d_a(x)\equiv\fft{\Delta_a}{2\pi}+x\pmod1,\qquad D_a(x)\equiv\left\lfloor\fft{\Delta_a}{2\pi}+x\right\rfloor.
\end{equation}
Note that, while $\mathcal G^{\mathrm{exp}}$ is a sum of twelve exponentially small terms, generically only a single one will dominate, depending on the relative magnitudes of $d_a(\sigma\hat k'/n')$.  

Given the asymptotic form of $\mathcal G(\hat j',\hat k')$, the $\tilde{\mathbb B}$ matrix can be expanded into the sum of an $\mathcal O(1)$ matrix and a subexponential one, $\tilde{\mathbb B}=\tilde{\mathbb B}^0+\tilde{\mathbb B}^{\mathrm{exp}}$.  If $\det\relax(1+\tilde{\mathbb B}^0)\ne0$, then we are essentially done, as it will not contribute at the $\mathcal O(1/\beta)$ order in the high-temperature limit.  However, if this vanishes, the subexponential contribution becomes important.  We thus consider the $\mathcal O(1)$ order determinant first, before turning to the subexponential one.

\subsubsection{$\mathcal O(1)$ order determinant}

For the $\tilde{\mathbb B}^0$ matrix, we note that its entries are built from $\mathcal G^0_{\{m',n',r'\}}(\hat j',\hat k')$, where here we have restored the sector labels $\{m',n',r'\}$.  However, examination of (\ref{G-fct:leading}) demonstrates that it is actually independent of $m'$ and $r'$ as well as the index $\hat j'$.  As a result, we can write the matrix expression
\begin{equation}
    \tilde{\mathbb B}^0_{\{m',n',r'\}}=\fft1{m'}U\otimes\tilde{\mathbb B}^0_{\{1,n',0\}},
\end{equation}
where $U$ is the $m'\times m'$ square matrix whose entries are all unity.  Since $U$ has only one non-vanishing eigenvalue equal to $m'$, we then see that
\begin{equation}
    \det(1+\tilde{\mathbb B}^0_{\{m',n',r'\}})=\det(1+\tilde{\mathbb B}^0_{\{1,n',0\}}),
\end{equation}
where the determinant on the left is that of an $N\times N$ matrix, while that on the right is of an $n'\times n'$ matrix.

At this point, we are still left with the $n'\times n'$ determinant to evaluate.  However, there is an important special case corresponding to $\{m',n',r'\}=\{N,1,0\}$.  This case is trivial since $\tilde{\mathbb B}^0_{\{1,1,0\}}=0$, so that
\begin{equation}
	\det(1+\tilde{\mathbb B}^0_{\{N,1,0\}})=1.
	\label{eq:N10}
\end{equation}
The situation is more complicated when $n'\ne1$.  While we do not have a proof, numerical evidence indicates that the $\mathcal O(1)$ order determinant only takes on two possibilities, depending on the chemical potentials:
\begin{equation}
    \det(1+\tilde{\mathbb B}^0_{\{m',n',r'\}})=0\mbox{ or }n'^2.
\end{equation}
In order to investigate where the determinant vanishes, we take $d_a=\Delta_a/2\pi\pmod1$ and assume none of them are integer multiples of $1/n'$ as in (\ref{eq:d:def}).  Furthermore, without loss of generality, we let $\sum_ad_a=1$, which follows from the requirement $\sum_a\Delta_a=0$. (The other possible case, $\sum_ad_a=2$, can be mapped to this one by taking into account the invariance of the index under $\Delta_a\to-\Delta_a$ discussed in (\ref{Deg:free}).)

\begin{figure}[t]
    \centering
    \includegraphics[width=.32\linewidth]{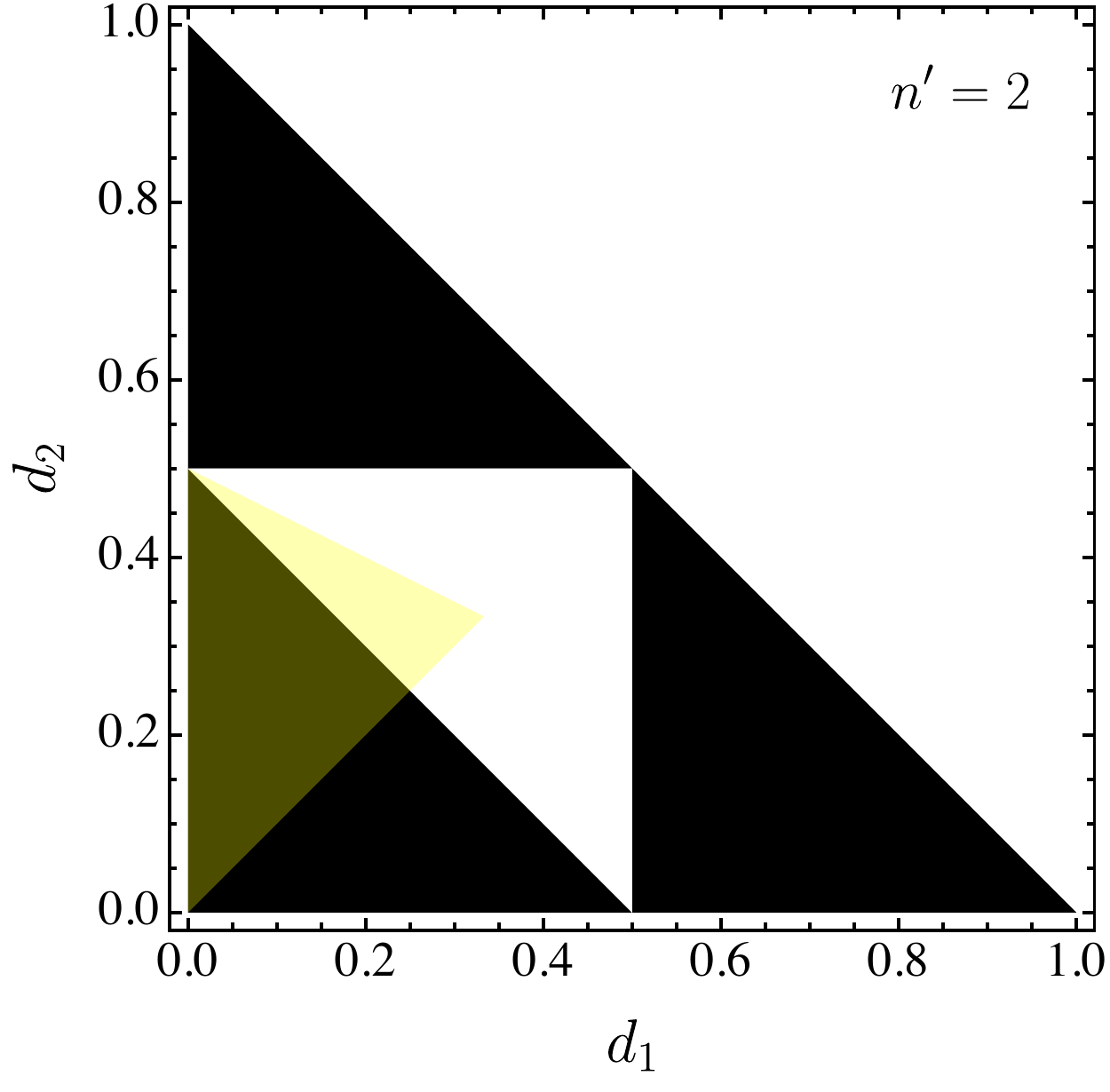}
    \includegraphics[width=.32\linewidth]{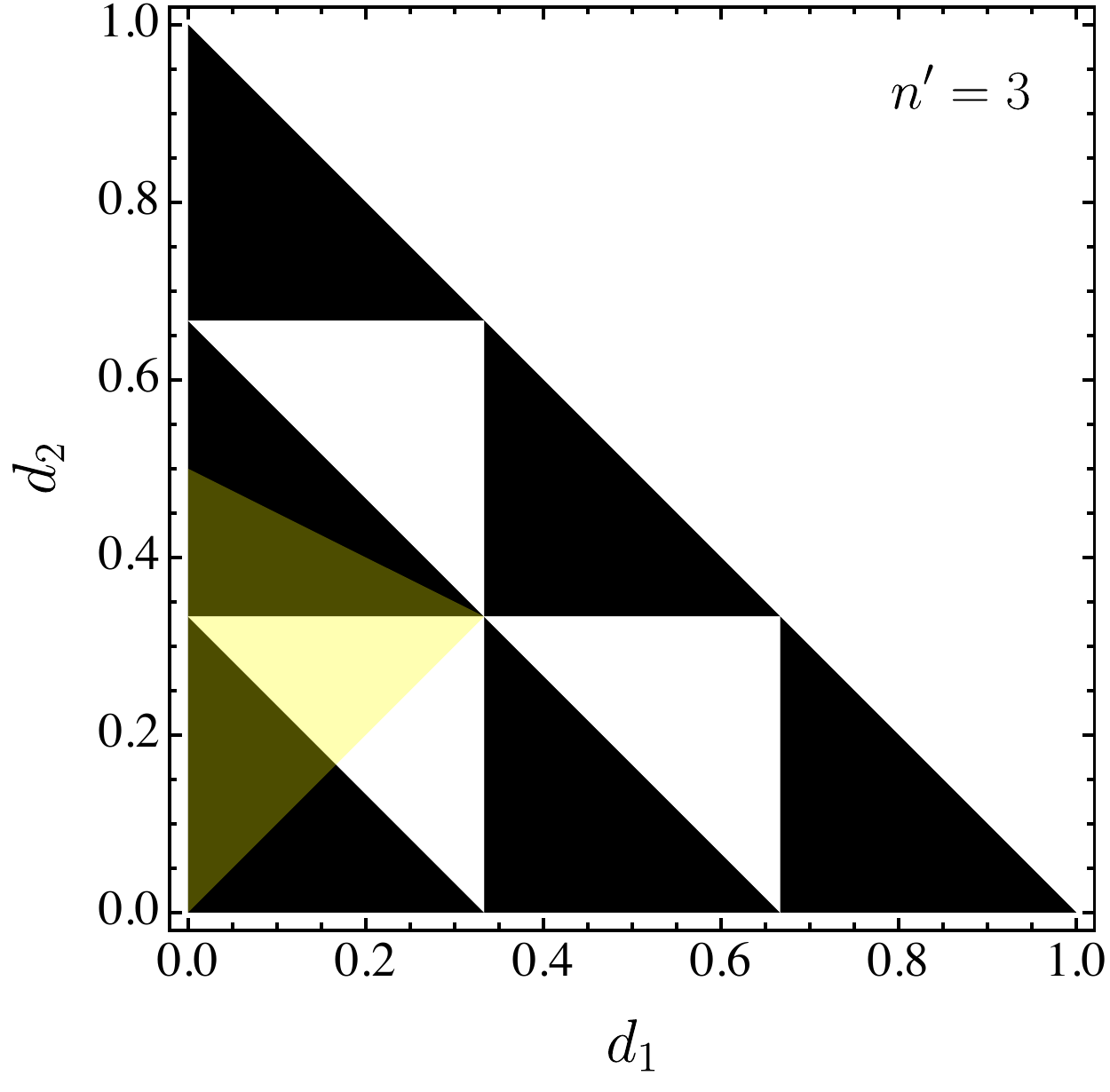}
    \includegraphics[width=.32\linewidth]{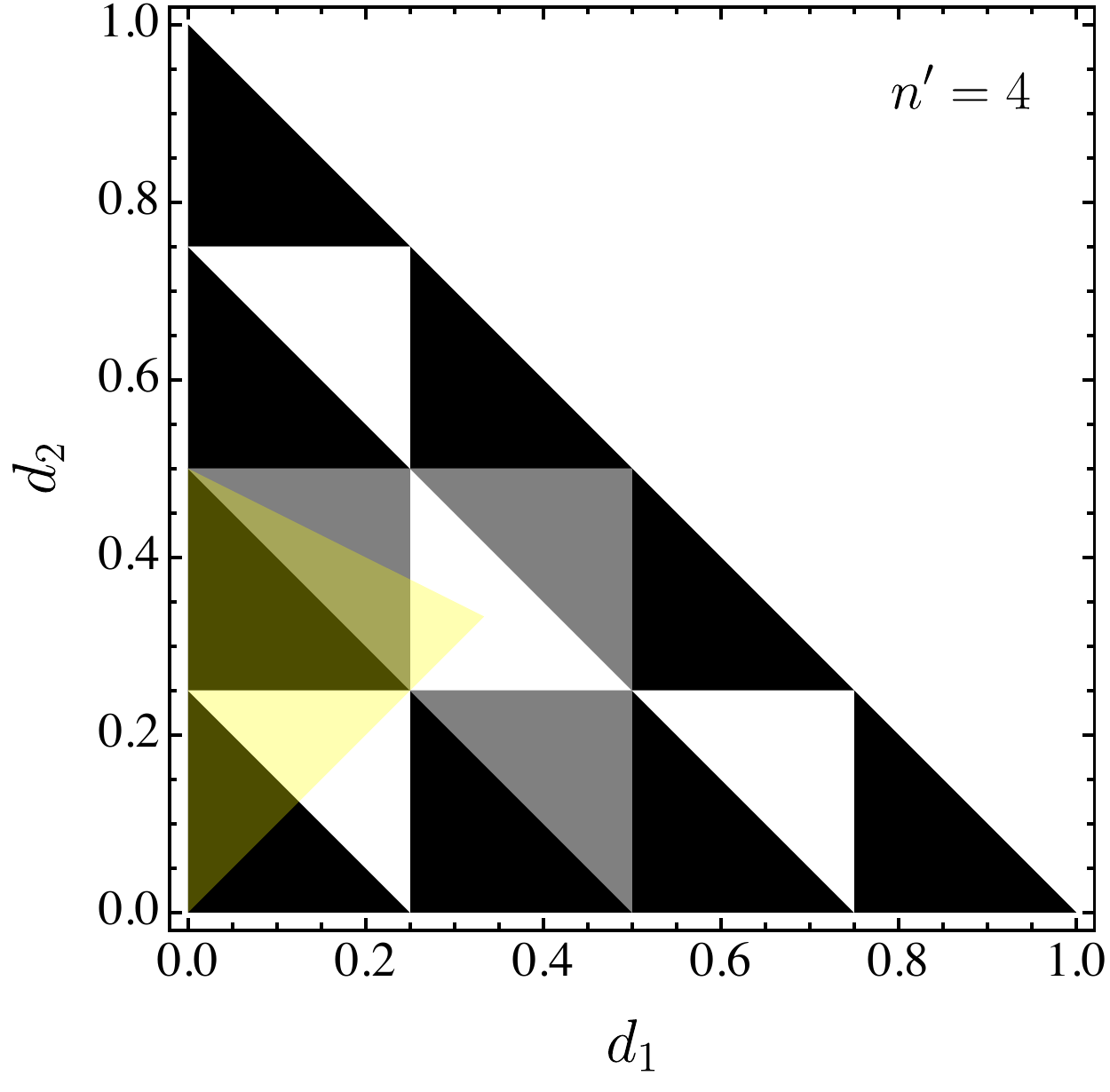}\\[1em]
    \includegraphics[width=.32\linewidth]{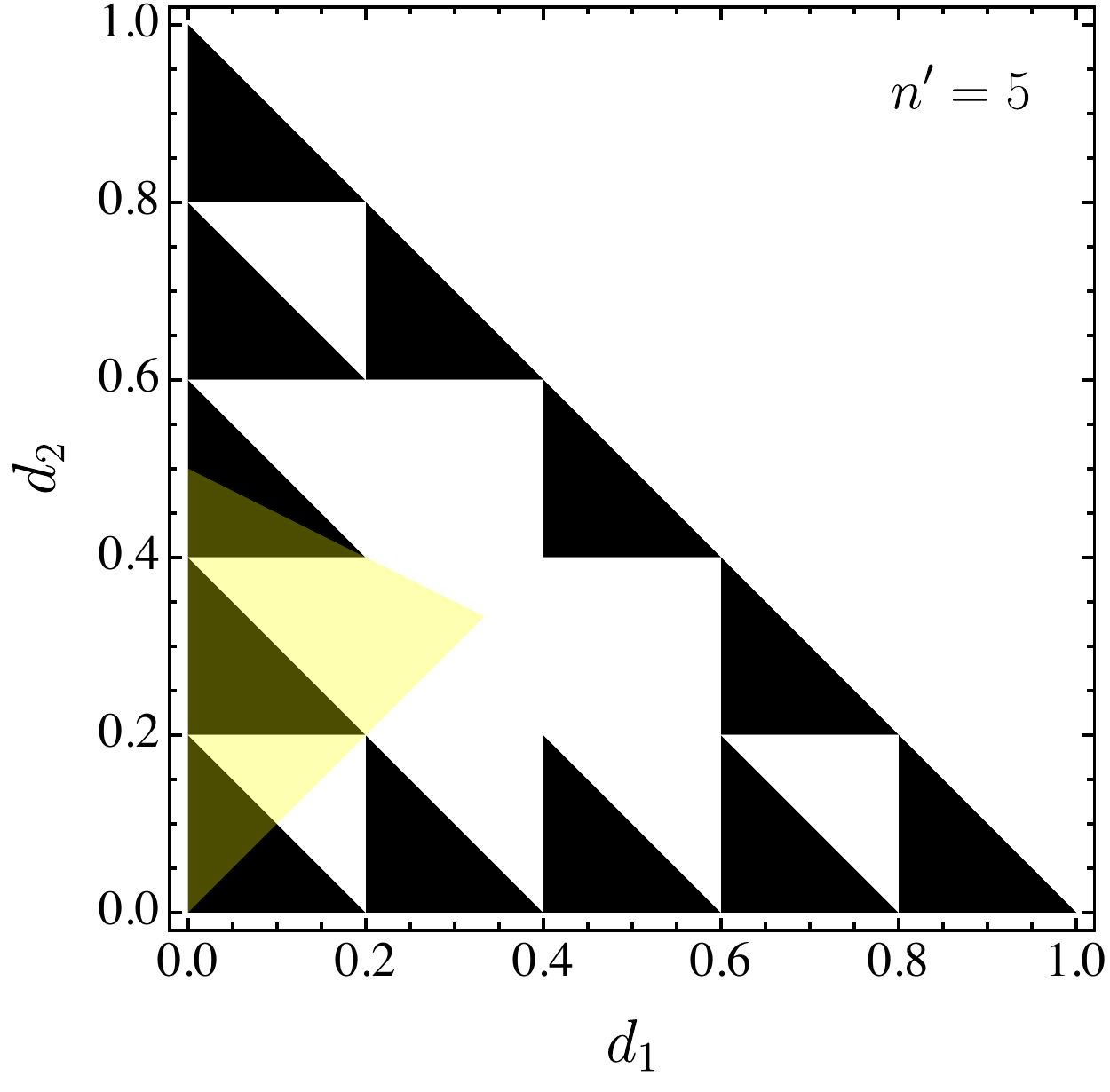}
    \includegraphics[width=.32\linewidth]{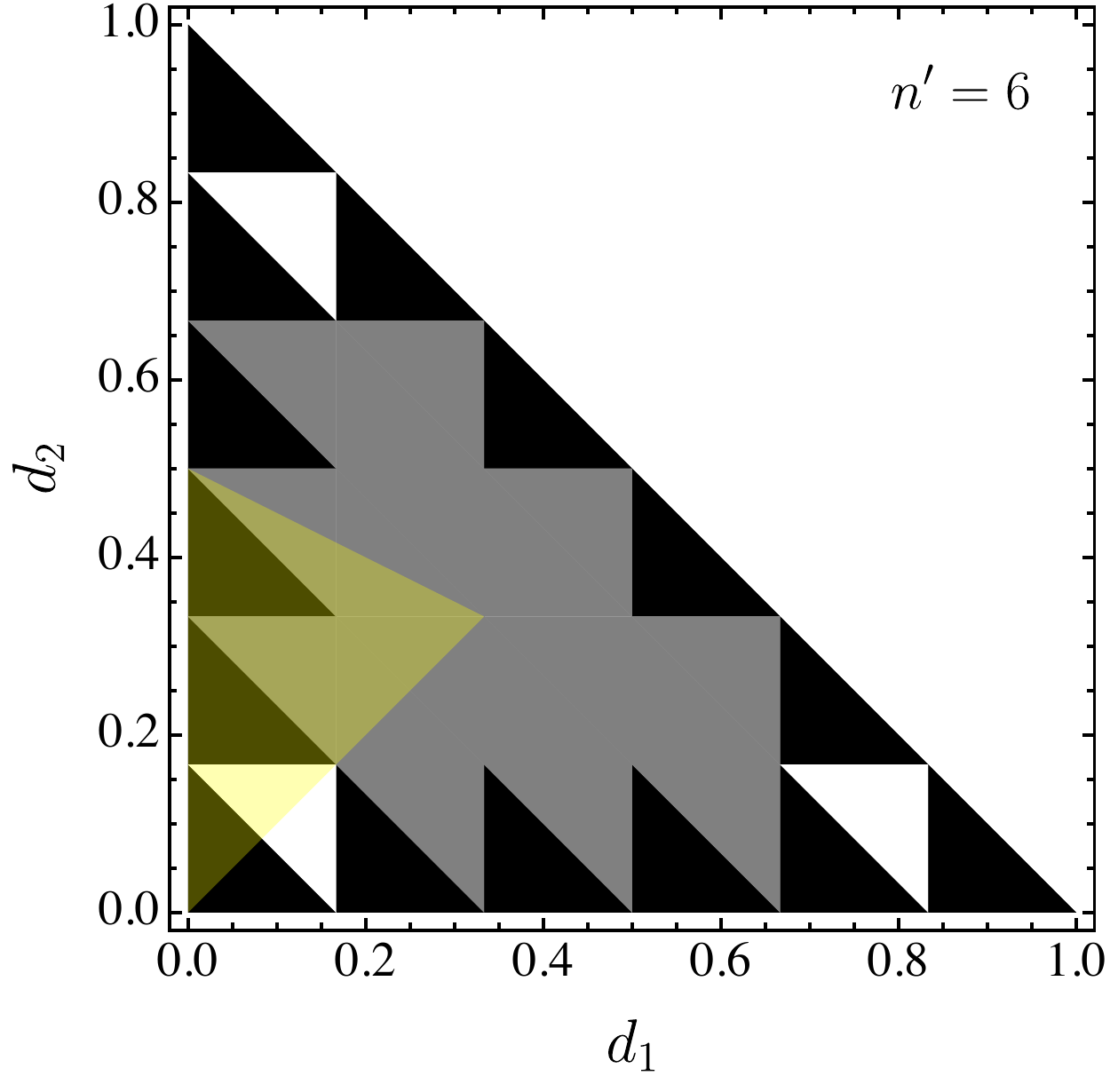}
    \includegraphics[width=.32\linewidth]{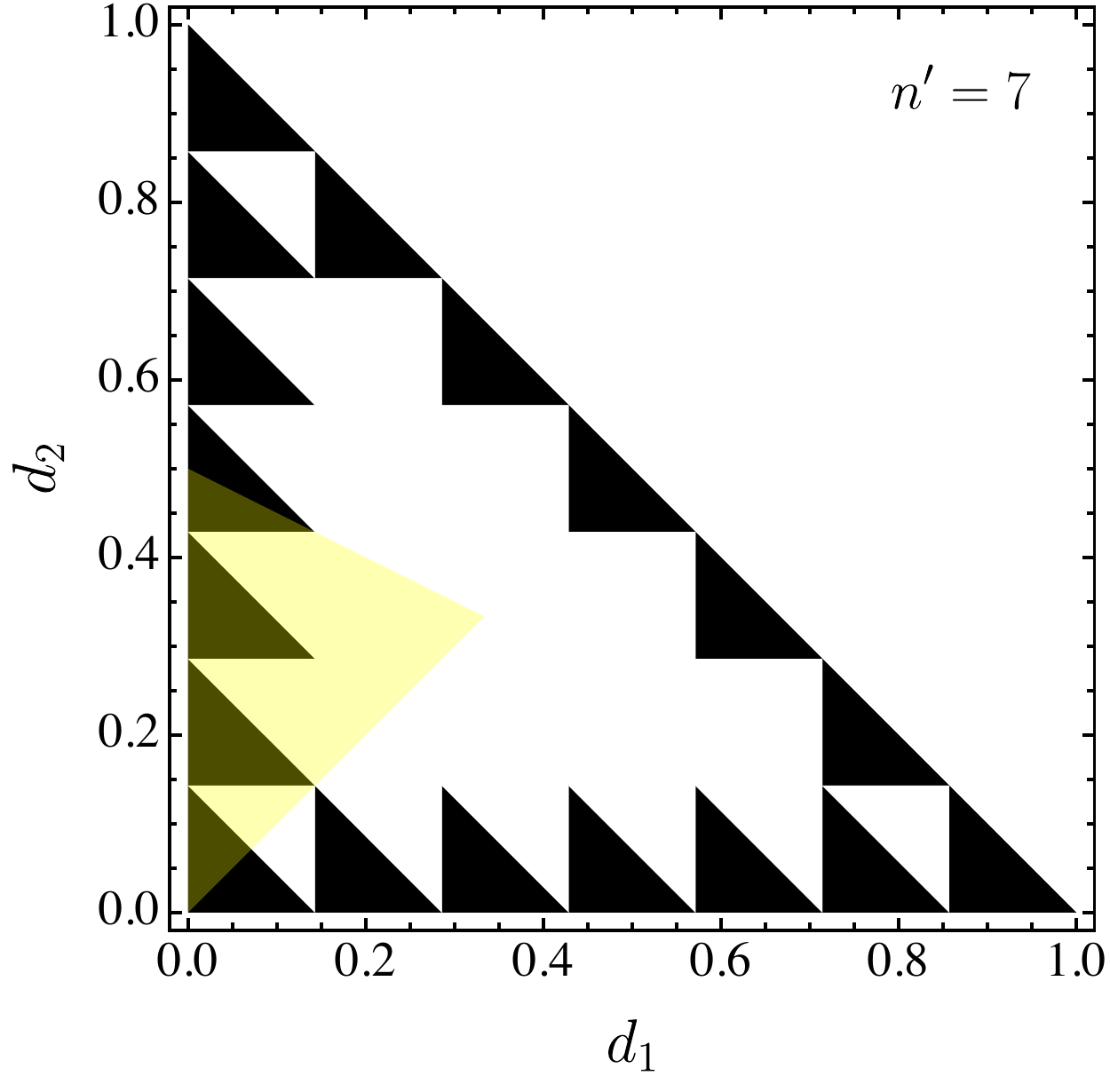}
    \caption{Regions of vanishing determinant for $n'=2,\ldots,7$.  The black regions correspond to $1+\tilde{\mathbb B}^0=0$, while the gray regions correspond to non-trivial $1+\tilde{\mathbb B}^0$, but still with vanishing determinant.  The determinant evaluates to $n'^2$ in the unshaded regions.  The yellow triange corresponds to the region $0<d_1\le d_2\le d_3<1$.}
    \label{fig:regions}
\end{figure}

For small values of $n'$, the regions in $(d_1,d_2)$ parameter space where the determinant vanishes are shown in Fig.~\ref{fig:regions}.  Here the $(n'-1)\times(n'-1)$ matrix $1+\tilde{\mathbb B}^0_{\{1,n',0\}}$ vanishes identically in the black regions. For prime $n'$, this appears to be the only places where the determinant vanishes, while for composite $n'$ there are additional regions with vanishing determinant but with non-trivial $1+\tilde{\mathbb B}^0$, represented by the gray regions.

For general $n'$, consider that the $\tilde{\mathbb B}^0$ matrix is obtained from
\begin{equation}
	\mathcal G^0(\hat j',\hat k')=\begin{cases}
	-1, & d_{a_2}>\mbox{min}(\hat k')\quad\mbox{and}\quad(d_{a_1}-\mbox{min}(\hat k'))(d_{a_3}-\mbox{max}(\hat k'))<0;\\
	0, & d_{a_1}<\mbox{min}(\hat k')\quad\mbox{and}\quad(d_{a_2}-\mbox{min}(\hat k'))(d_{a_3}-\mbox{max}(\hat k'))<0;\\
	1, & d_{a_2}<\mbox{min}(\hat k')<d_{a_3}<\mbox{max}(\hat k');\\
	2, & d_{a_3}<\mbox{min}(\hat k'),
	\end{cases}
\end{equation}
which is a direct consequence of (\ref{G-fct:leading}). Here $d_a$'s are ordered as $0<d_{a_1}\leq d_{a_2}\leq d_{a_3}<1$ and we have defined min($\hat k'$) and max($\hat k'$) as the min and max of $\{\hat k'/n',1-\hat k'/n'\}$, respectively. In particular, note that $\mathcal G^0(\hat j',\hat k')=-\delta_{0,\hat k'}$ where
\begin{align}
	\fft{l'}{n'}<d_{a_2}< d_{a_1}+d_{a_2}<\fft{l'+1}{n'},\quad\mbox{with}\quad l'=0,1,\cdots,\left\lfloor\fft{n'-1}{2}\right\rfloor,
	\label{lead:vanish}
\end{align}
which corresponds to the black regions in Fig.~\ref{fig:regions}. Inserting $\mathcal G^0(\hat j',\hat k')=-\delta_{0,\hat k'}$ into (\ref{Btilde:entry}) then explains why $1+\tilde{\mathbb B}^0$ vanishes identically in these regions. However, the resulting $\tilde{\mathbb B}^0$ matrix outside of the black regions is rather difficult to work with. Nevertheless, for prime $n'$, we conjecture based on numerical evidence that $\det\relax(1+\tilde{\mathbb B}^0)=n'^2$ everywhere outside of the black regions specified by (\ref{lead:vanish}). The case for composite $n'$ is clearly more complicated, as can be seen from the figure.

\subsubsection{First subexponential order determinant}

Whenever the $\mathcal O(1)$ order determinant vanishes, it becomes necessary to examine the exponentially suppressed contributions to $\tilde{\mathbb B}$.  For prime $n'$, we can derive
\begin{equation}
	1+\tilde{\mathbb B}_{\{1,n',r'\}}=
	\exp\left[2\pi i\left(\fft{\gamma_{\{1,n',r'\}}(d_a)}{n'}r'-\fft{\alpha_{\{1,n',r'\}}(d_a)}{\tau}\right)\right]C(\gamma_{\{1,n',r'\}}(d_a)),
\end{equation}
at leading order whenever we are in the black regions specified by (\ref{lead:vanish}). Here $C(\gamma)$ is an $(n'-1)\times(n'-1)$-square matrix defined by
\begin{equation}
	[C(\gamma)]_{\mu,\nu}=2\delta_{\mu,\nu}+\delta_{\mu,\gamma}+\delta_{\mu,n'-\gamma}-\delta_{\mu-\nu,\gamma}-\delta_{\mu-\nu,n'-\gamma}-\delta_{\mu-\nu,-\gamma}-\delta_{\mu-\nu,-(n'-\gamma)},
\end{equation}
and $\alpha_{\{1,n',r'\}}(d_a)$ and $\gamma_{\{1,n',r'\}}(d_a)$ are given by
\begin{subequations}
	\begin{align}
	\alpha_{\{1,n',r'\}}(d_a)&=\begin{cases}
	d_{a_3}-\fft{n'-1}{n'}, & l'=0;\\
	\min\left\{d_{a_2}-\fft{l'}{n'},d_{a_3}-\fft{n'-l'-1}{n'}\right\}, & l'\neq 0,
	\end{cases}\\
	\gamma_{\{1,n',r'\}}(d_a)&=\begin{cases}
	n'-l', & \alpha_{\{1,n',r'\}}(d_a)=d_{a_2}-\fft{l'}{n'};\\
	l'+1, & \alpha_{\{1,n',r'\}}(d_a)=d_{a_3}-\fft{n'-l'-1}{n'}.
	\end{cases}
	\end{align}
\end{subequations}
Note that here we are excluding the special case $d_{a_2}-\fft{l'}{n'}=d_{a_3}-\fft{n'-l'-1}{n'}$. 

For prime $n'$, we can prove $\det C(\gamma)=n'^2$ for any $\gamma=1,\ldots,n'-1$.  In particular, we can first show $\det C(1)=n'^2$ by mathematical induction. Then since
\begin{equation}
	C(\gamma)=\tilde\sigma_\gamma^{-1}C(1)\tilde\sigma_\gamma,
\end{equation}
with a permutation $\sigma_\gamma(\mu)=\gamma\mu\pmod {n'}$ and the corresponding permutation matrix
\begin{equation}
	[\tilde\sigma_\gamma]_{\mu,\nu}=\delta_{\sigma_\gamma(\mu),\nu},
\end{equation}
we have $\det C(\gamma)=n'^2$ for any $\gamma=1,\cdots,n'-1$.  Combining this result with the conjecture for the $\mathcal O(1)$ behavior made above, we find for prime $n'$ (excluding the special case)
\begin{equation}
	\det(1+\tilde{\mathbb B}_{\{1,n',r'\}})=n'^2
	\exp\left[2\pi i(n'-1)\left(\fft{\gamma_{\{1,n',r'\}}(d_a)}{n'}r'-\fft{\alpha_{\{1,n',r'\}}(d_a)}{\tau}\right)\right],
	\label{observ:4}
\end{equation}
at leading order, where we set $\alpha_{\{1,n',r'\}}(d_a)$ and $\gamma_{\{1,n',r'\}}(d_a)$ to vanish outside of the black regions specified by (\ref{lead:vanish}). 

We now have all the components needed to work out the high temperature expansion of the index in the $\{m,n,r\}$ sector provided the corresponding $m'$ is unity and $n'$ is prime.  For such a sector, substituting (\ref{observ:4}) into (\ref{Index:highT:mnr}) yields
\begin{align}
	\log Z_{\{m,n,r\}}(\tau;\Delta_a,\mathfrak n_a)&=\fft{2\pi^2}{\beta}\biggl[\sum_{a=1}^3(1-\mathfrak n_a)\left(d_a(1-d_a)-m'^2x_a(1-x_a)\right)\nn\\
	&\kern3.5em+2(n'-1)\alpha_{\{m',n',r'\}}(d_a)\biggr]\nn\\
	&\quad-3\log n'+i\varphi+\mathcal O(e^{-c_{\{m',n',r'\}}(d_a)/\beta}),
	\label{eq:flogZ}
\end{align}
where $\varphi$ is a phase independent of $\tau$ and $c_{\{1,n',r'\}}(d_a)>0$ is a positive function away from special values of the chemical potentials $d_a$. We expect that this expression continues to hold for arbitrary values of $\{m,n,r\}$, although we have been unable to obtain a general expression for the determinant factor $\alpha_{\{m',n',r'\}}(d_a)$ apart from the above case.

\subsubsection{The $N=2$ and $3$ cases}
\label{Small:N:eg:1}

We now give a couple of examples supporting the results (\ref{observ:4}) and (\ref{eq:flogZ}).  For notational convenience, here we set $0<d_1\leq d_2\leq d_3<1$ without loss of generality and therefore the domain in $(d_1,d_2)$ parameter space shrinks down to the yellow triangle in Fig.~\ref{fig:regions}. 

For the $N=2$ case, we have a total of three sectors, labeled by $\{m',n',r'\}=\{2,1,0\}$, $\{1,2,0\}$ and $\{1,2,1\}$.  The determinant in the $\{2,1,0\}$ sector is trivial as seen in (\ref{eq:N10}), so we focus on the $\{1,2,r'\}$ case. From (\ref{Btilde:entry}) and (\ref{G-fct:leading}), we have
\begin{equation}
	\det(1+\tilde{\mathbb B}_{\{1,2,r'\}})=\fft{4\sum_{a=1}^3D_{a}(1/2)-2\mathcal G_{\{1,2,r'\}}^{\mathrm{exp}}(0,1)}{1+2\sum_{a=1}^3D_{a}(1/2)-\mathcal G_{\{1,2,r'\}}^{\mathrm{exp}}(0,1)-\mathcal G_{\{1,2,r'\}}^{\mathrm{exp}}(0,0)},
	\label{det:N=2}
\end{equation}
up to higher order terms. Due to the constraint $\sum_aD_a=-1$, the sum $\sum_a D_{a}(1/2)$ is restricted as
\begin{equation}
	\sum_{a=1}^3 D_a(1/2)=\begin{cases}
	-1, & d_3<1/2;\\
	0, & d_3>1/2,
	\end{cases}
\end{equation}
and therefore (\ref{det:N=2}) leads to (``$\sim$'' denotes the non-vanishing leading order)
\begin{align}
	\det(1+\tilde{\mathbb B}_{\{1,2,r'\}})\sim\begin{cases}
	4, & d_3<1/2;\\
	-2\mathcal G^{\mathrm{exp}}_{\{1,2,r'\}}(0,1), & d_3>1/2.
	\end{cases}
\end{align}
When $d_3>1/2$, we use the expansion (\ref{G-fct:exp}) to obtain
\begin{align}
	\mathcal G_{\{1,2,r'\}}^{\mathrm{exp}}(0,1)\sim-2\exp\left[2\pi i\left(\fft{r'}{2}-\fft{d_3-1/2}{\tau}\right)\right]\qquad(d_3>1/2),
\end{align}
where we used the fact that
\begin{equation}
	\min\{d_{a}(1/2),1-d_{a}(1/2)\}=d_{3}(1/2)=d_3-\fft{1}{2},
\end{equation}
which is valid for $d_3>1/2$. Consequently, we have
\begin{align}
	\det(1+\tilde{\mathbb B}_{\{1,2,r'\}})\sim4\times\begin{cases}
	1, & d_3<1/2;\\
	(-1)^{r'}\exp\left(-2\pi i\fft{d_3-1/2}{\tau}\right), & d_3>1/2,
	\end{cases}
\end{align}
which is consistent with (\ref{observ:4}).  As a result, the $N=2$ index is given by (\ref{eq:flogZ}) with
\begin{equation}
    \alpha_{\{2,1,0\}}=0,\qquad\alpha_{\{1,2,0\}}=\alpha_{\{1,2,1\}}=\max(0,d_3-1/2).
\end{equation}

We now turn to the $N=3$ index.  Here there are four sectors, given by $\{m',n',r'\}=\{3,1,0\}$ and $\{1,3,r'\}$ with $r'=0,1,2$.  Since the determinant in the $\{3,1,0\}$ sector is trivial as seen in (\ref{eq:N10}), we focus on the $\{1,3,r'\}$ case. From (\ref{Btilde:entry}) and (\ref{G-fct:leading}), we have
\begin{align}
	\det(1+\tilde{\mathbb B}_{\{1,3,r'\}})=&\left(\fft{3\sum_a(D_a(1/3)+D_a(2/3))-2\mathcal G^{\mathrm{exp}}(0,1)-\mathcal G^{\mathrm{exp}}(0,2)}{1+2\sum_a(D_a(1/3)+D_a(2/3))-\mathcal G^{\mathrm{exp}}(0,0)-\mathcal G^{\mathrm{exp}}(0,1)-\mathcal G^{\mathrm{exp}}(0,2)}\right)\nn\\
	&\times\Bigl(\mathcal G^{\mathrm{exp}}(0,1)\leftrightarrow\mathcal G^{\mathrm{exp}}(0,2)\Bigr),
\end{align}
up to higher order terms, where we have suppressed the $\{1,3,r'\}$ subscript from $\mathcal G(\hat j',\hat k')$. Due to the constraint $\sum_aD_a=-1$, the sum $\sum_a(D_a(1/3)+D_a(2/3))$ is restricted as
\begin{equation}
	\sum_{a=1}^3(D_a(1/3)+D_a(2/3))=\begin{cases}
	-1, & d_2<1/3<d_3<2/3;\\
	0, & \mbox{otherwise},
	\end{cases}
\end{equation}
and therefore we have
\begin{equation}
	\det(1+\tilde{\mathbb B}_{\{1,3,r'\}})\sim\begin{cases}
	9, &\kern-3em d_2<1/3<d_3<2/3;\\
	2\mathcal G^{\mathrm{exp}}(0,1)^2+5\mathcal G^{\mathrm{exp}}(0,1)\mathcal G^{\mathrm{exp}}(0,2)+2\mathcal G^{\mathrm{exp}}(0,2)^2, & \mbox{otherwise}.
	\end{cases}
\end{equation}
We can pull out the leading order behavior of $\mathcal G^{\mathrm{exp}}(0,\hat k')$ with $\hat k'=1,2$ from (\ref{G-fct:exp}).  The result is independent of $\hat k'$, and is given by
\begin{align}
	\mathcal G^{\mathrm{exp}}_{\{1,3,r'\}}(0,\hat k')\sim\begin{cases}
	-\exp\left[2\pi i\left(\fft{r'}{3}-\fft{d_3-2/3}{\tau}\right)\right], & d_2<\fft{1}{3},\,d_3>\fft{2}{3};\\-\exp\left[2\pi i\left(\fft{2r'}{3}-\fft{d_2-1/3}{\tau}\right)\right], & d_2>\fft{1}{3},\,d_3<\fft{2}{3},
	\end{cases}
\end{align}
where we made use of 
\begin{align}
	&\min\{d_{a}(1/3),1-d_{a}(1/3),d_{a}(-1/3),1-d_{a}(-1/3)\}\nn\\
	&\kern12em=\begin{cases}
	d_{3}(1/3)=d_3-\fft{2}{3}, & d_2<1/3,\,d_3>2/3;\\
	d_{2}(-1/3)=d_2-\fft{1}{3}, & d_2>1/3,\,d_3<2/3.
	\end{cases}
\end{align}
Consequently, we have
\begin{equation}
	\det(1+\tilde{\mathbb B}_{\{1,3,r'\}})\sim9\times\begin{cases}
	1, & d_2<1/3,\,d_3<2/3;\\
	\exp\left[4\pi i\left(\fft{r'}{3}-\fft{d_3-2/3}{\tau}\right)\right], & d_2<1/3,\,d_3>2/3;\\
	\exp\left[4\pi i\left(\fft{2r'}{3}-\fft{d_2-1/3}{\tau}\right)\right], & d_2>1/3,\,d_3<2/3,
	\end{cases}
\end{equation}
which is consistent with (\ref{observ:4}).  As a result, the $N=3$ index is given by (\ref{eq:flogZ}) with
\begin{equation}
    \alpha_{\{3,1,0\}}=0,\qquad\alpha_{\{1,3,r'\}}=\max(0,d_2-1/3,d_3-2/3).
\end{equation}
%

\subsection{The full index in the `high-temperature' limit}

After the above examination of the individual $\{m,n,r\}$ sectors, we now return to the full index, (\ref{eq:Ztotsum}), in the high temperature limit.  From (\ref{eq:flogZ}), we expect the leading behavior of each individual sector $Z_{\{m,n,r\}}$ to scale exponentially in $1/\beta$.  Thus, the sectors with the largest positive coefficient of $1/\beta$ will dominate the full index, and the other sectors will be exponentially suppressed.  As a result, we are left with identifying the dominant sectors and their contribution to the index. Note that the degeneracy, if any, of the dominant sectors does not contribute to the leading order expansion of the full index.

The high temperature limit was investigated in \cite{Hosseini:2016cyf}, where the BAE were solved in the $\{1,N,0\}$ sector.  Substituting the corresponding $\{m',n',r'\}=\{N,1,0\}$ into (\ref{eq:flogZ}) gives
\begin{equation}
    \log Z_{\{1,N,0\}}=-\fft{2\pi^2}\beta(N^2-1)\sum_a(1-\mathfrak n_a)d_a(1-d_a)+\mathcal O(1).
\end{equation}
As discussed in \cite{Hosseini:2016cyf}, this is to be extremized with respect to the potentials, $d_a$ under the constraint $\sum_{a=1}^3 d_a=1$.  This can be performed by the method of Lagrange multipliers, and the result is
\begin{equation}
	\left.\log Z_{\{1,N,0\}}\right|_{\bar d_a}=\frac{\pi^2}{6\beta}c_r(\mathfrak n_a)+\cdots,
	\label{Hosseini}
\end{equation}
where the extremum values, $\bar d_a$, are given by
\begin{equation}\label{Extremum:1N0}
	\bar d_a=\fft{\mathfrak n_a(\mathfrak n_a-1)}{2\Theta},\qquad
	\Theta\equiv1-(\mathfrak n_1\mathfrak n_2+\mathfrak n_2\mathfrak n_3+\mathfrak n_3\mathfrak n_1),
\end{equation}
and $c_r(\mathfrak n_a)$ is the right-moving central charge of the 2d $\mathcal N=(0,2)$ SCFT arising from the KK compactification of the topologically twisted $\mathcal N=4$ SYM over $S^2$ \cite{Hosseini:2016cyf}. Here we have assumed that two of $\mathfrak n_a$'s are negative so $c_r(\mathfrak n_a)$ given in (\ref{central:charge}) is positive.

Note that the left-hand side of (\ref{Extremum:1N0}) only corresponds to a single sector of the full index.  Nevertheless, this connection to the right-moving central charge suggests that the $\{1,N,0\}$ sector is a dominant one, so that
\begin{equation}\label{Claim:highT:result}
	\log Z\big|_{\bar d_a}=\frac{\pi^2}{6\beta}c_r(\mathfrak n_a)+\cdots,
\end{equation}
where $Z$ is the full index, and is indeed the only physically relevant quantity to connect to the central charge.  Note that, if $\log Z$ is truly dominated by $\log Z_{\{1,N,0\}}$, then $\bar d_a$ can be considered not just an extremum of $\log Z_{\{1,N,0\}}$, but the full index as well.  Hence the identification of the central charge with the extremized index in the `high-temperature' limit \cite{Hosseini:2016cyf} remains valid in the presence of multiple BAE solutions.

Of course, it is still necessary to demonstrate that the $\{1,N,0\}$ sector is a dominant one.  To do so, we must show that
\begin{equation}
	\left.\Re\log\left(\fft{Z_{\{1,N,0\}}}{Z_{\{m,n,r\}}}\right)\right|_{\bar d_a}\geq 0,
\end{equation}
at leading order in $1/\beta$ for any $\{m,n,r\}$. This inequality can be written explicitly by inserting (\ref{Extremum:1N0}) into (\ref{eq:flogZ}):
\begin{equation}\label{Claim:highT:explicit}
	\sum_{a=1}^3(1-\mathfrak n_a)\left(\fft{\bar x_a(1-\bar x_a)}{n'^2}-\bar d_a(1-\bar d_a)\right)\ge \fft{2(n'-1)}{N^2}\alpha_{\{m',n',r'\}}(\bar d_a),
\end{equation}
for any $\{m',n',r'\}$.  The difficulty in proving this inequality lies in the $\alpha_{\{m',n',r'\}}(\bar d_a)$ factors which originate from the determinant of $1+\tilde{\mathbb B}$.

\subsubsection{The case of vanishing $\alpha_{\{m',n',r'\}}(\bar d_a)$}

As we have noted, the $\alpha_{\{m',n',r'\}}(\bar d_a)$ factors depend in a complicated manner on the extremized potentials $\bar d_a$.  However, they are always non-negative and in fact vanish in the white regions of Fig.~\ref{fig:regions}.  In this case the claim (\ref{Claim:highT:explicit}) reduces to
\begin{equation}
	\sum_{a=1}^3(1-\mathfrak n_a)\left(\fft{\bar x_a(1-\bar x_a)}{n'^2}-\bar d_a(1-\bar d_a)\right)\geq 0,
	\label{eq:reduced}
\end{equation}
for any integers $\mathfrak n_a$ with the constraints $\sum_a\mathfrak n_a=2$ and two of them being negative. Note that the latter is necessary for the 2d SCFT arising from the KK compactification to have a positive right-moving central charge \cite{Benini:2013cda}. Here we prove this reduced claim under the same constraints, but without the integer condition. To begin with, note that the map (\ref{Extremum:1N0}) is in fact invertible between
\begin{equation}
	\{\mathfrak n_a:\sum_a\mathfrak n_a=2,~\mbox{two of them are negative}\}~~\leftrightarrow~~\{\bar d_a:\sum_a\bar d_a=1,~(1/2-\bar d_1-\bar d_2)^2>\bar d_1\bar d_2\},
	\label{domain}
\end{equation}
with the inverse map
\begin{equation}
	\mathfrak n_a=\fft{2\bar d_a(2\bar d_a-1)}{1-4(\bar d_1\bar d_2+\bar d_2\bar d_3+\bar d_3\bar d_1)}.
\end{equation}
Hence, using 
\begin{equation}
	(1-\mathfrak n_a)(1-2\bar d_a)=-\fft{\prod_{a=1}^3(1-\mathfrak n_a)}{\Theta}>0,
\end{equation}
the above claim can be rewritten equivalently as
\begin{equation}
	\sum_{a=1}^3\fft{1}{1-2\bar d_a}\left(\fft{\bar x_a(1-\bar x_a)}{n'^2}-\bar d_a(1-\bar d_a)\right)\geq 0,
	\label{Claim:w/o}
\end{equation}
for any $\bar d_a$ within the domain given in (\ref{domain}). Now we define $f(\bar d_1,\bar d_2)$ as the LHS of the above inequality. Then within the subdomain of fixed $\lfloor n'\bar d_a\rfloor$, where $\partial_{\bar d_a}f$ is well defined, we can consider the extremum of $f$ which satisfies
\begin{align}
	&\partial_{\bar d_1}f=\partial_{\bar d_2}f=0\nn\\&\Rightarrow~\fft{2}{(1-\bar d_a)^2}\left(\fft{\bar x_a(1-\bar x_a)}{n'^2}-\bar d_a(1-\bar d_a)\right)+\fft{1}{1-2\bar d_a}\left(\fft{1-2\bar x_a}{n'}-(1-2\bar d_a)\right)=k,
\end{align}
where $k$ is some constant independent of $a$. At this extremum, the determinant of the Hessian is given by
\begin{equation}
	\begin{vmatrix}
	\partial_{\bar d_1}^2f & \partial_{\bar d_1}\partial_{\bar d_2}f\\ \partial_{\bar d_2}\partial_{\bar d_1}f & \partial_{\bar d_2}^2f
	\end{vmatrix}=\fft{16k^2}{(1-2\bar d_1)(1-2\bar d_2)(1-2\bar d_3)}<0,
\end{equation}
so it is in fact a saddle point. Note that we have used $\bar d_{a_1}\leq\bar d_{a_2}<1/2<\bar d_{a_3}$, ordered as before, which is valid in the domain given in (\ref{domain}). This implies the minimum of $f$ within the subdomain of fixed $\lfloor n'\bar d_a\rfloor$ must stay on its boundary. If one investigates the values of $f$ on this boundary, it is straightforward (though tedious) to check that $f$ is minimized where
\begin{align}
	\bar x_{a_1}\to 0^+,\bar x_{a_2}\to 0^+,\bar x_{a_3}\to 1^-\quad&\mbox{for}\quad\sum_{a=1}^3\lfloor n'\bar d_a\rfloor=n'-1,\\
	\bar x_{a_1}\to 0^+,\bar x_{a_2}\to 1^-,\bar x_{a_3}\to 1^-\quad&\mbox{for}\quad\sum_{a=1}^3\lfloor n'\bar d_a\rfloor=n'-2.
\end{align}
For both cases, we have
\begin{equation}
	f\to\sum_{a=1}^3\fft{-\bar d_a(1-\bar d_a)}{1-2\bar d_a}=-\fft{1}{4}\prod_{a=1}^3\fft{\mathfrak n_a}{1-\mathfrak n_a}\geq 0,
\end{equation}
which proves (\ref{Claim:w/o})  and thereby the claim (\ref{Claim:highT:explicit}) in the white regions of Fig.~\ref{fig:regions} where $\alpha_{\{m',n',r'\}}(\bar d_a)$ vanishes.

\subsubsection{The $N=2$ and $3$ cases}
\label{Small:N:eg:2}

Of course, we are left to deal with the regions where $\alpha_{\{m',n',r'\}}(\bar d_a)$ is strictly positive.  In this case, the inequality (\ref{Claim:highT:explicit}) is stronger than the reduced claim (\ref{eq:reduced}), and the above proof no longer applies.  In the absence of a general expression for $\alpha_{\{m',n',r'\}}(\bar d_a)$, we only 
verify (\ref{Claim:highT:explicit}) for $N=2$ and $3$, and leave the general case for $N\ge4$ as a conjecture. Here we set $0<\bar d_1\leq\bar d_2\leq\bar d_3<1$ without loss of generality as in \ref{Small:N:eg:1}.

For $N=2$, it suffices to prove the inequality (\ref{Claim:highT:explicit}) for $\{m',n',r'\}=\{1,2,r'\}$. Inserting $\{\bar x_1,\bar x_2,\bar x_3\}=\{2\bar d_1,2\bar d_2,2\bar d_3-1\}$ and $\alpha_{\{1,2,r'\}}(\bar d_a)=\bar d_3-1/2$ into (\ref{Claim:highT:explicit}) then reduces the claim to
\begin{equation}
	\fft{(\mathfrak n_3-1)(\mathfrak n_1\mathfrak n_2-1)}{\Theta}\geq 0,
\end{equation}
which is true since $\mathfrak n_2\leq\mathfrak n_1\leq-1$ and $\mathfrak n_3\geq4$. Hence the claim is proven for $N=2$.

For $N=3$, it suffices to prove the inequality (\ref{Claim:highT:explicit}) for $\{m',n',r'\}=\{1,3,r'\}$. Inserting 
\begin{subequations}
\begin{align}
	\{\bar x_1,\bar x_2,\bar x_3\}&=\begin{cases}
	\{3\bar d_1,3\bar d_2,3\bar d_a-1\},& (\bar d_2<1/3,\,\bar d_3<2/3);\\
    \{3\bar d_1,3\bar d_2,3\bar d_a-2\},& (\bar d_2<1/3,\,\bar d_3>2/3);\\
	\{3\bar d_1,3\bar d_2-1,3\bar d_a-1\},& (\bar d_2>1/3,\,\bar d_3<2/3),
	\end{cases}\\
	\alpha_{\{1,3,r'\}}(\bar d_a)&=\begin{cases}
	\bar d_3-\fft{2}{3}, & (\bar d_2<1/3,\,\bar d_3>2/3);\\
	\bar d_2-\fft{1}{3}, & (\bar d_2>1/3,\,\bar d_3<2/3),
	\end{cases}
\end{align}
\end{subequations}
into (\ref{Claim:highT:explicit}) and examining the resulting expression then proves the claim for $N=3$.

\section{Discussion}
\label{sec:5}

Our main observation is that the BAEs for the topologically twisted index for $\mathcal N=4$, $SU(N)$ SYM on $T^2\times S^2$ have multiple solutions labeled by three integers $m$, $n$, and $r$ such that $N=mn$ and $r=0,\ldots,n-1$.  Modular covariance of the index is only achieved after summing over a complete set of these solutions.  Taking this into account, we verified that the index gives the elliptic genus of the $(0,2)$ theory \cite{Closset:2013sxa,Honda:2015yha}, which transforms as a weak Jacobi form of weight zero.  Based on this observation, we expect that the BAEs for general supersymmetric indices where there is a $T^2$ factor will similarly admit multiple solutions.  This is equivalent to having multiple saddle points in the matrix integrals that arise from localization of the path integral.

Multiple solutions of the BAEs, however, make it rather difficult to compute the index explicitly. This is because we have to sum over all possible contributions to get the full index (\ref{eq:Ztotsum}). We conjecture that the contribution from a single sector, namely $Z_{\{1,N,0\}}$, will dominate in the `high-temperature' limit when extremized with respect to the flavor chemical potentials $\Delta_a$, giving the result (\ref{Claim:highT:result}), which connects the index to the central charge of the (0,2) theory.  However, we have been unable to demonstrate this in full generality because of the difficulty in computing $\det\mathbb B_{\{m',n',r'\}}$ in this limit.  

This connection between the high-temperature limit of the topologically twisted index and the right-moving central charge, (\ref{central:charge}), was derived with the assumption that $c_r(\mathfrak n_a)>0$.  On the holographic side, positivity of the central charge is necessary for a good AdS$_3\times S^2$ supergravity solution to exist.  However, it may be interesting to explore the case when a single magnetic charge $\mathfrak n_a$ is negative, corresponding to $c_r(\mathfrak n_a)<0$ after extremization.  While the holographic dual is not obviously well-defined, the field theory may still be interesting on its own.  In this situation, the $\{1,N,0\}$ sector may no longer dominate, and additional sectors will have to be considered as well.

We were initially drawn to the topologically twisted $T^2\times S^2$ index because of our interest in its large-$N$ limit.  This limit, however, is somewhat delicate, as the sum over sectors involves the modular parameter $\tilde\tau=(m\tau+r)/n$ with $N=mn$.  The different sectors then have $\Im\tilde\tau$ ranging from $(\Im\tau)/N\to0$ to $N\Im\tau\to\infty$ for fixed $\tau$ in the large-$N$ limit.  Similar to the high-temperature limit, we may expect $\mathcal O(N^2)$ contributions to arise from the $\Im\tilde\tau\to0$ sectors, and in particular the $\{1,N,0\}$ sector.  However, for finite modular parameter $\tau$, the final result ought to remain a weak Jacobi form of weight zero, as the Cardy limit would not yet have been taken.

Assuming progress can be made with the large-$N$ limit, this would allow us to investigate the partition function for microstate counting of the dual magnetic black string, in analogy with the AdS$_4$ black hole story of \cite{Benini:2015eyy}. However, an analytic supergravity solution has not yet been constructed. (See \cite{Chamseddine:1999xk} for a singular magnetic string and \cite{Benini:2013cda} for a numerical solution.) So in order to complete the picture, it would be worth obtaining such a solution that interpolates from an AdS$_3\times S^2$ near-horizon geometry \cite{Benini:2013cda} to asymptotic AdS$_5$ with conformal boundary $T^2\times S^2$.  If such an analytic solution can be found, an interesting follow up would be to compare the $\log N$ term in the index with the corresponding one-loop supergravity result. (See \cite{Liu:2017vll,Jeon:2017aif,Liu:2017vbl} for recent work on the topologically twisted index for ABJM theory.) This would, however, require a more careful computation of $\log\det\mathbb B$ than what we considered above, and hence may remain an open challenge.

\acknowledgments

The motivation to explore the large-$N$ limit of the topologically twisted index on $T^2\times S^2$ arose out of conversations with L.\ Pando Zayas.  We wish to thank S.\ M.\ Hosseini, F.\ Larsen, L.\ Pando Zayas and V.\ Rathee for enlightening discussions and N.\ Bobev for interesting comments. This work was supported in part by the US Department of Energy under Grant No.\ DE-SC0007859.

\appendix

\section{Elliptic functions}\label{App A}

Let $q=e^{2\pi i\tau}$ and $x=e^{iu}$.  Then the Dedekind eta function is given by
\begin{equation}\label{eq:eta}
	\eta(q)=\eta(\tau)=q^{\fft1{24}}\prod_{n=1}^\infty(1-q^n)
	=\sum_{n=-\infty}^\infty(-1)^nq^{\fft32(n-\fft16)^2}.
\end{equation}
The Jacobi theta function $\theta_1$ is given by
\begin{align}
	\theta_1(x;q)=\theta_1(u;\tau)&=-iq^{\fft18}(x^{\fft12}-x^{-\fft12})\prod_{n=1}^\infty(1-q^n)(1-xq^n)(1-x^{-1}q^n)\nn\\
	&=-i\sum_{n=-\infty}^\infty(-1)^n x^{n+\frac{1}{2}}q^{\fft12(n+\frac{1}{2})^2}.
	\label{eq:theta1}
\end{align}

These elliptic functions satisfy the following modular properties
\begin{subequations}
\begin{align}
	\eta(\tau+1)&=e^{i\pi/12}\eta(\tau),&\eta(-1/\tau)&=\sqrt{-i\tau}\eta(\tau),\label{modular of eta}\\
	\theta_1(u;\tau+1)&=e^{i\pi/4}\theta_1(u;\tau),&\theta_1(u/\tau;-1/\tau)&=-i\sqrt{-i\tau}e^{iu^2/4\pi\tau}\theta_1(u;\tau).\label{modular of theta}
\end{align}
\label{modular}
\end{subequations}
These modular properties, (\ref{modular}), can be extended to general $SL(2;\mathbb Z)$ transformations
\begin{subequations}
\begin{align}
	\eta\left(\fft{a\tau+b}{c\tau+d}\right)&=\xi\sqrt{c\tau+d}\,\eta(\tau),\\
	\theta_1\left(\fft{u}{c\tau+d},\fft{a\tau+b}{c\tau+d}\right)&=\xi^3\sqrt{c\tau+d}\,e^{\fft{icu^2}{4\pi(c\tau+d)}}\theta_1(u,\tau),
\end{align}
\label{eq:sl2z}
\end{subequations}
where $\xi$ is a 24-th root of unity.

In addition, $\theta_1$ is quasi-doubly periodic with ($p,q\in\mathbb Z$)
\begin{equation}
	\theta_1(u+2\pi(p+q\tau);\tau)=(-1)^{p+q}e^{-iq u}e^{-i\pi q^2\tau}\theta_1(u;\tau).\label{eq:uperiod}
\end{equation}

In the text, we have introduced the weak Jacobi form of weight $-1$ and index $1/2$,
\begin{equation}
    \psi(u;\tau)\equiv\fft{\theta_1(u;\tau)}{\eta(\tau)^3}.
\end{equation}
(This is the square-root of the unique weak Jacobi form of weight $-2$ and integer index $1$, sometimes denoted $\varphi_{-2,1}$.)  This can be expanded for $\Im\tau\gg1$ (ie $|q|\ll1$), with the result
\begin{align}
    \psi(u;\tau)&\sim i(-1)^\ell e^{\fft{u_2^2}{4\pi\tau_2}}|q|^{-\fft12\delta(1-\delta)}e^{i\left(\pi\ell(\ell+1)\tau_1-(\ell+\fft12)u_1\right)}\nn\\
    &\qquad\times\left(1-e^{-i(2\pi\ell\tau_1-u_1)}|q|^\delta-e^{i(2\pi(\ell+1)\tau_1-u_1)}|q|^{1-\delta}+\mathcal O(|q|)\right),
\end{align}
where
\begin{equation}
    \delta=\fft{u_2}{2\pi\tau_2}\pmod 1,\qquad\ell=\left\lfloor\fft{u_2}{2\pi\tau_2}\right\rfloor.
\end{equation}
Note that this expansion breaks down for integer $\ell$. We can also rewrite this expansion as
\begin{align}
	\psi(u;\tau)&\sim -i(-1)^{\ell'}e^{\fft{u_2^2}{4\pi\tau_2}}|q|^{-\fft12\delta'(1-\delta')}e^{i\left(\pi\ell'(\ell'+1)\tau_1+(\ell'+\fft12)u_1\right)}\nn\\
	&\qquad\times\left(1-e^{-i(2\pi\ell'\tau_1+u_1)}|q|^{\delta'}-e^{i(2\pi(\ell'+1)\tau_1+u_1)}|q|^{1-\delta'}+\mathcal O(|q|)\right),
	\label{psi:highT}
\end{align}
where
\begin{equation}
	\delta'=-\fft{u_2}{2\pi\tau_2}\pmod 1=1-\delta,\qquad\ell'=\left\lfloor-\fft{u_2}{2\pi\tau_2}\right\rfloor=-\ell-1.
\end{equation}
%

\section{Invariance of $\det\mathbb B$ under $T$ and $S$ transformations}
\label{App E}

Here we demonstrate that $\det\mathbb B$ transforms according to (\ref{T:inv:condi2}) and (\ref{S:inv:condi2}) under $T$ and $S$ transformations, respectively.  We first note that the eigenvalues for the BAE solution denoted by $\{m,n,r\}$ are canonically ordered according to (\ref{Index:pair}). The key step here is then to order the eigenvalues for the BAE solution denoted by $\{m',n',r'\}$ differently, according to 
\begin{subequations}
\begin{align}
	\{m,n,r\}~\mbox{sector}&:~u_{n\hat j+\hat k}\quad\to\quad u_{\hat j\hat k},\\
	\{m',n',r'\}~\mbox{sector}&:~u_{n\hat j+\hat k}\quad\to\quad u_{\hat j'\hat k'}.
\end{align}
\end{subequations}
Note that $(\hat j,\hat k)\to(\hat j',\hat k')$ is a bijective map from $\mathbb Z_m\times\mathbb Z_n$ to $\mathbb Z_{m'}\times\mathbb Z_{n'}$ for both $T$ and $S$ transformation cases so the above ordering for $\{m',n',r'\}$ sector is valid. Furthermore, it does not affect the determinant of the $\mathbb B$ matrix as the determinant does not depend on eigenvalue ordering.

Now we prove, with respect to the above ordering,
\begin{subequations}
\begin{align}
	\mathbb B_{\{m,n,r\}}(\Delta_a;\tau+1)&=\mathbb B_{\{m',n',r'\}}(\Delta_a;\tau),\\
	\mathbb B_{\{m,n,r\}}(\Delta_a/\tau;-1/\tau)&=\tau^{N-1}\mathbb B_{\{m',n',r'\}}(\Delta_a;\tau),
\end{align}
\end{subequations}
which automatically yields (\ref{T:inv:condi2}) and (\ref{S:inv:condi2}) respectively. Note that $\{m',n',r'\}$ are different for $T$ and $S$ cases. From (\ref{N B matrix}), the $(l,N)$ entries of the LHS and the RHS are the same as unity for $l\in\{1,\cdots,N\}$. In order to prove that the remaining entries also match, it suffices to show
\begin{subequations}
\begin{align}
	&\mathcal G_{\{m,n,r\}}(\hat j-\hat j_0,\hat k-\hat k_0;\Delta_a,\tau+1)=\mathcal G_{\{m',n',r'\}}(\hat j'-\hat j_0',\hat k'-\hat k_0';\Delta_a,\tau),\label{T:inv:proof}\\
	&\mathcal G_{\{m,n,r\}}(\hat j-\hat j_0,\hat k-\hat k_0;\Delta_a/\tau,-1/\tau)=\tau\mathcal G_{\{m',n',r'\}}(\hat j'-\hat j_0',\hat k'-\hat k_0';\Delta_a,\tau),\label{S:inv:proof}
\end{align}
\end{subequations}
for any $\hat j,\hat j_0\in\mathbb Z_m$ and $\hat k,\hat k_0\in\mathbb Z_n$. Note that these are not trivial from (\ref{G-fct:T}) or (\ref{G-fct:S}) but can be proved based on those relations and the following properties of the $\mathcal G$-function:
\begin{subequations}
\begin{align}
	&\mathcal G_{\{m,n,r\}}(\hat j+m,\hat k;\Delta_a,\tau)=\mathcal G_{\{m,n,r\}}(\hat j,\hat k;\Delta_a,\tau),\\
	&\mathcal G_{\{m,n,r\}}(\hat j,\hat k+n;\Delta_a,\tau)=\mathcal G_{\{m,n,r\}}(\hat j+r,\hat k;\Delta_a,\tau).
\end{align}\label{G-fct:period}
\end{subequations}

\subsection*{Proof of (\ref{T:inv:proof})}
\begin{align}
	LHS&=\mathcal G_{\{m,n,r\}}\left(\left\{\hat j-\hat j_0+r\left\lfloor\fft{\hat k-\hat k_0}{n}\right\rfloor,m\right\},\{\hat k-\hat k_0,n\};\Delta_a,\tau+1\right)\nonumber\\
	&=\mathcal G_{\{m',n',r'\}}\left(\left\{\hat j-\hat j_0+r\left\lfloor\fft{\hat k-\hat k_0}{n}\right\rfloor+\{\hat k-\hat k_0,n\}\left\lfloor\fft{m+r}{n}\right\rfloor,m\right\},\{\hat k-\hat k_0,n\};\Delta_a,\tau\right)\nonumber\\
	&=\mathcal G_{\{m',n',r'\}}\left(\left\{\hat j+\hat k\left\lfloor\fft{m+r}{n}\right\rfloor,m\right\}-\left\{\hat j_0+\hat k_0\left\lfloor\fft{m+r}{n}\right\rfloor,m\right\},\hat k-\hat k_0;\Delta_a,\tau\right)\nn\\
	&=RHS
\end{align}
Here $\{A,B\}$ denotes $A$ mod $B$ ($0\leq A<B$).
Note that (\ref{G-fct:period}) has been used in the 1st and the 3rd lines. The 2nd line comes from (\ref{G-fct:T}).

\subsection*{Proof of (\ref{S:inv:proof})}
\begin{align}
	LHS&=\mathcal G_{\{m,n,r\}}\left(\left\{\hat j-\hat j_0+r\left\lfloor\fft{\hat k-\hat k_0}{n}\right\rfloor,m\right\},\{\hat k-\hat k_0,n\};\fft{\Delta_a}{\tau},-\fft{1}{\tau}\right)\nonumber\\&=\tau\mathcal G_{\{m',n',r'\}}\Bigg(\left\{-\fft{g}{n}\left((\hat k-\hat k_0)+d\left\{\fft{n}{g}(\hat j-\hat j_0)+\fft{r}{g}(\hat k-\hat k_0),\fft{N}{g}\right\}\right),g\right\},\nonumber\\
	&\kern7em\left\{\fft{n}{g}(\hat j-\hat j_0)+\fft{r}{g}(\hat k-\hat k_0),\fft{N}{g}\right\};\Delta_a,\tau\Bigg)\nonumber\\
	&=\tau\mathcal G_{\{m',n',r'\}}\Bigg(\left\{-\fft{g}{n}\left(\hat k+d\left\{\fft{n}{g}\hat j+\fft{r}{g}\hat k,\fft{N}{g}\right\}\right),g\right\}\nn\\
	&\kern7em-\left\{-\fft{g}{n}\left(\hat k_0+d\left\{\fft{n}{g}\hat j_0+\fft{r}{g}\hat k_0,\fft{N}{g}\right\}\right),g\right\},\nonumber\\
	&\kern7em\left\{\fft{n}{g}\hat j+\fft{r}{g}\hat k,\fft{N}{g}\right\}-\left\{\fft{n}{g}\hat j_0+\fft{r}{g}\hat k_0,\fft{N}{g}\right\};\Delta_a,\tau\Bigg)=RHS.
\end{align}
Note that (\ref{G-fct:period}) has been used in the 1st and the 4th lines. The 2nd line comes from (\ref{G-fct:T}) followed by the identity $M\{A,B\}=\{MA,MB\}$.

\section{Proof that the map (\ref{S:inv:G}) is bijective}\label{App D}

First we prove that (\ref{S:inv:G}) is one-to-one, i.e.
\begin{equation}
	\hat j'_1=\hat j'_2~\&~\hat k'_1=\hat k'_2\quad\Rightarrow\quad\hat j_1=\hat j_2~\&~\hat k_1=\hat k_2.
\end{equation}
To begin with, note that (\ref{S:inv:G:a}) implies
\begin{equation}
	\hat j'_1=\hat j'_2~\&~\hat k'_1=\hat k'_2\quad\Rightarrow\quad\hat k_1=\hat k_2\pmod n,
\end{equation}
which means $\hat k_1=\hat k_2$ in fact. Combined with this fact, (\ref{S:inv:G:b}) implies 
\begin{equation}
	\hat j'_1=\hat j'_2~\&~\hat k'_1=\hat k'_2\quad\Rightarrow\quad\hat j_1=\hat j_2\pmod m,
\end{equation}
which means $\hat j_1=\hat j_2$ in fact. Hence (\ref{S:inv:G}) is one-to one.

Next we prove that (\ref{S:inv:G}) is onto, i.e.\ there exists $(\hat j,\hat k)\in\mathbb Z_m\times\mathbb Z_n$ satisfying (\ref{S:inv:G}) for any given $(\hat j',\hat k')\in\mathbb Z_{m'}\times\mathbb Z_{n'}$ $(m'=g,~n'=N/g)$. To begin with, recall that we have
\begin{equation}
	\fft{n}{g}(-b)+\fft{r}{g}(d)=1.
\end{equation}
Then for any given $(\hat j',\hat k')\in\mathbb Z_{m'}\times\mathbb Z_{n'}$, we have
\begin{equation}
	\fft{n}{g}\left(-b\hat k'+\fft{r}{g}\hat j'\right)+\fft{r}{g}\left(d\hat k'-\fft{n}{g}\hat j'\right)=\hat k'.
\end{equation}
This can be rewritten as
\begin{equation}
	\hat k'=\left\{\fft{n}{g}\left\{-b\hat k'+\fft{r}{g}\hat j'+r\left\lfloor\fft{d\hat k'-\fft{n}{g}\hat j'}{n}\right\rfloor,m\right\}+\fft{r}{g}\left\{d\hat k'-\fft{n}{g}\hat j',n\right\},\fft{N}{g}\right\}.
\end{equation}
As in Appendix~\ref{App E}, $\{A,B\}$ denotes $A$ mod $B$ ($0\leq A<B$). Now it is straightforward to check that
\begin{subequations}
\begin{align}
	\hat j&=\left\{-b\hat k'+\fft{r}{g}\hat j'+r\left\lfloor\fft{d\hat k'-\fft{n}{g}\hat j'}{n}\right\rfloor,m\right\}\in\mathbb Z_m,\\
	\hat k&=\left\{d\hat k'-\fft{n}{g}\hat j',n\right\}\in\mathbb Z_n,
\end{align}
\end{subequations}
truly satisfy (\ref{S:inv:G}), so (\ref{S:inv:G}) is onto.

\bibliographystyle{JHEP}
\bibliography{indexrefs}





\end{document}